\RequirePackage{filecontents}

\RequirePackage{snapshot}
\documentclass[aps,prd,reprint,showpacs,secnumarabic,groupedaddress,amsmath,amssymb]{preprint}

\renewcommand\documentclass[2][]{}
\let\origparagraph=\paragraph
\AtBeginDocument{%
  \let\paragraph=\origparagraph
}
\makeatother
\pdfoutput=1

\documentclass[%
  aps,
  prd,
  twocolumn,
  groupedaddress,
  showpacs,
  floatfix,
  amsmath,
  amssymb
]{revtex4-1}

\usepackage[T1]{fontenc}
\usepackage{txfonts}
\usepackage{microtype}
\usepackage{graphicx}
\graphicspath{{figs/}}
\usepackage{color}
\usepackage{bm}

\usepackage[breaklinks=true,
pdfauthor={Anton W{\"o}llert, Michael Klaiber, Heiko Bauke and Christoph H. Keitel},
pdftitle={Relativistic tunneling picture of electron-positron pair creation}]{hyperref}

\definecolor{darkblue}{RGB}{0,0,90}
\hypersetup{
  pdfborder={0 0 0},
  colorlinks=true,
  linkcolor=darkblue,
  anchorcolor=darkblue,
  citecolor=darkblue,
  filecolor=darkblue,
  menucolor=darkblue,
  runcolor=darkblue,
  urlcolor=darkblue,
}

\newcommand{\ket}[1]{%
\mathchoice{%
\left|#1\right\rangle}{%
|#1\rangle}{%
\left|#1\right\rangle}{%
\left|#1\right\rangle}%
}
\newcommand{\bra}[1]{%
\mathchoice{%
\left \langle #1 \right|}{%
\langle #1 |}{%
\left \langle #1 \right|}{%
\left \langle #1 \right|}%
}
\newcommand{\scalprod}[2]{%
\mathchoice{%
\left\langle #1 \vert #2 \right\rangle}{%
\langle #1 | #2 \rangle}{%
\left\langle #1 \vert #2 \right\rangle}{%
\left\langle #1 \vert #2 \right\rangle}%
}

\newcommand{\op}[1]{\hat{#1}}

\newcommand{\vect}[1]{\boldsymbol{#1}}

\newcommand{\norm}[1]{\vert #1 \vert}

\DeclareMathOperator{\sgn}{sgn}
\DeclareMathOperator{\arcosh}{arcosh}

\renewcommand{\Im}[1]{\operatorname{Im}{#1}}

\newcommand{\dd}{\mathrm{d}}

\begin{document}

\clubpenalty = 1000
\widowpenalty = 1000 
\displaywidowpenalty = 1000

\title{Relativistic tunneling picture of electron-positron pair creation}

\author{Anton W{\"o}llert}%
\email{woellert@mpi-hd.mpg.de}
\affiliation{Max-Planck-Institut f{\"u}r Kernphysik, Saupfercheckweg 1, 69117 Heidelberg, Germany}

\author{Michael Klaiber}
\affiliation{Max-Planck-Institut f{\"u}r Kernphysik, Saupfercheckweg 1, 69117 Heidelberg, Germany}

\author{Heiko Bauke}
\email{heiko.bauke@mpi-hd.mpg.de}
\affiliation{Max-Planck-Institut f{\"u}r Kernphysik, Saupfercheckweg 1, 69117 Heidelberg, Germany}

\author{Christoph H.~Keitel}
\affiliation{Max-Planck-Institut f{\"u}r Kernphysik, Saupfercheckweg 1, 69117 Heidelberg, Germany}

\date{\today}

\begin{abstract}
  The common tunneling picture of electron-positron pair creation in a
  strong electric field is generalized to pair creation in combined
  crossed electric and magnetic fields.  This enhanced picture, being
  symmetric for electrons and positrons, is formulated in a
  gauge-invariant and Lorentz-invariant manner for quasistatic fields.
  It may be used to infer qualitative features of the pair creation
  process.  In particular, it allows for an intuitive interpretation of
  how the presence of a magnetic field modifies and, in particular
  cases, even enhances pair creation.  The creation of electrons and
  positrons from the vacuum may be assisted by an energetic photon,
  which can also be incorporated into this picture of pair creation.
\end{abstract}


\maketitle

\section{Introduction and motivation}
\label{sec:intro}

One of the most intriguing predictions of quantum electrodynamics (QED)
is certainly the possible breakdown of the vacuum in the presence of
ultrastrong electromagnetic fields into pairs of electrons and
positrons.  Since its first prediction by Sauter and others
\cite{sauter1931verhalten,heisenberg1936folgerungen,schwinger1951on_gauge,brezin1970pair},
pair creation has been studied in many papers, see
Refs.~\cite{ehlotzky2009fundamental,ruffini2010electron,dipiazza2012extremely}
for recent reviews.  The Schwinger critical field strength of
$E_\text{S}=1.3\times10^{18}\,\text{V/m}$, where spontaneous pair
creation is expected to set in, cannot be reached even by the strongest
lasers available today.  However, pair creation may be assisted by
additional fields or particles.  Current research covers, among others,
pair creation in spatially and temporally oscillating electric fields
\cite{gies2005pair,ruf2009pair,mocken2010nonperturbative,hebenstreit2011particle},
pair creation induced by the interaction of strong pulsed laser fields
with relativistic electron beams or a nuclear Coulomb field
(Bethe-Heitler process)
\cite{sokolov2010pair,mueller2012polarization,augustin2013bichromatic,krajewska2013betheheitler},
and pair creation induced by additional photons in the presence of an
external field
\cite{schuetzhold2008dynamically,dunne2009catalysis,piazza2009barrier,king2012thermal,krajewska2012breit}.
Furthermore, different aspects of pair creation like the effect of
magnetic fields
\cite{kim2006schwinger,su2012magnetic,su2012suppression}, the dynamics,
real-time evolution and pair distributions with nontrivial field
configurations in one effective dimension
\cite{lv2013pairtunnel,hebenstreit2013one,liu2014population,hebenstreit2014massless,jiang2014pair},
and effective mass signatures in multiphoton pair creation
\cite{kohlfurst2014effective} are investigated at present.  Current
interest in the old topic of pair creation is prompted by recent
advances in laser technology \cite{mourou2006optics,mourou2012exawatt}
aiming for laser intensities exceeding $10^{22}\,\mathrm{W/cm^2}$
(corresponding to electric field strengths of about
$10^{14}\,\mathrm{V/m}$) and by experimental proposals for quantum
simulators \cite{szpak2012optical_lattice} that may allow to study pair
creation via quantum simulation.  The ultimate quest for higher and
higher laser intensities for studying quantum electrodynamic effects
such as pair creation may be limited, however, just by the onset of pair
creation \cite{fedotov2010limitations}.

At field strengths below the Schwinger limit $E_\mathrm{S}$, pair
creation via ultrastrong electric fields may be interpreted as a
tunneling effect \cite{popov1972pair} similar to tunnel ionization from
bound states via strong electric fields
\cite{keldysh1965ionization,popov1973imaginary_time} using the method of
imaginary times \cite{popov2005imaginary} (see also recent applications
in Refs.~\cite{popru2008strong,popru2009multi,cortes2011laser}).  This
method uses classical trajectories with imaginary times to approximate
the exponential suppression for the transition amplitude of interest.  In
spite of conceptual difficulties
\cite{reiss2008limits,reiss2014tunneling}, the tunneling picture of
ionization was recently extended into the relativistic domain, where the
laser's magnetic field component can no longer be neglected
\cite{klaiber2013under,yakaboylu2013relativistic}.  The purpose of this
contribution is to establish a similar picture for pair creation in an
electromagnetic field, including the magnetic field to full extent.  The
influence of an additional quantized photon is incorporated and special
care is taken with respect to gauge and Lorentz invariance.  The
tunneling picture presented here is established in the quasistatic
limit, where the electromagnetic field is assumed to be constant and
uniform during the pair creation process.  Furthermore, spin effects are
not taken into account.

The manuscript is organized as follows: In Sec.~\ref{sec:theory} we
describe the electromagnetic field configuration for pair creation, lay
down the theoretical framework, and introduce all necessary notations.
Our main results are presented in Sec.~\ref{sec:tunneling_picture},
where the tunneling picture for electron-positron pair creation in
quasistatic crossed electric and magnetic fields is developed.  Three
different cases need to be distinguished, depending on the electric
field amplitude being larger than, equal to, or smaller than the
magnetic field amplitude.  The effect of a quantized photon is also
discussed for all these cases.  Properties of the maximum probability
trajectories, stemming from the imaginary time method, are investigated
in Sec.~\ref{sec:max_prob_traj}.  Finally, we conclude in
Sec.~\ref{sec:conclusions}.  Further details of the calculations have
been deferred into two appendixes.

\section{Theoretical framework}
\label{sec:theory}

\subsection{Geometric setup and notation}

\begin{figure}[b]
  \centering
  \includegraphics[width=\linewidth]{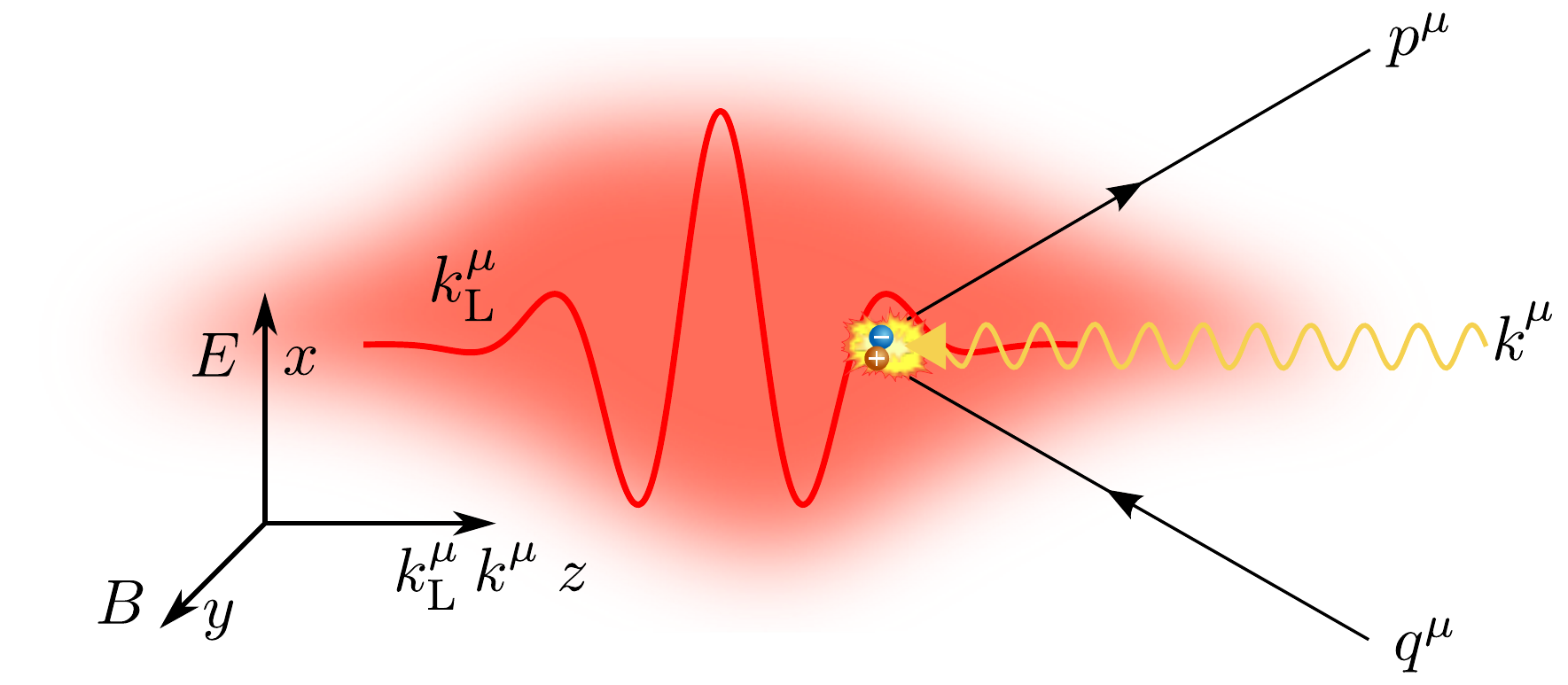}
  \caption{(color online) Schematic illustration, representing the
    geometry of the physical setup.  An electron and a positron with the
    kinetic four-momenta $p^\mu$ and $q^\mu$ are created in the presence
    of a strong external electromagnetic field (red shading), e.\,g., a
    plane wave or two colliding laser pulses.  Pair production may be
    assisted by an additional high-energy photon with four-momentum
    $k^\mu$ (yellow).  The directions of the electric field, the magnetic
    field, and the photon momentum are perpendicular to each other and
    parallel to the coordinate axes.}
  \label{fig:geometry}
\end{figure}

Natural units will be used in this work, i.\,e., $c = \hbar = 1$.  The
setup of the considered pair production process in an electromagnetic
field of an ultrastrong laser is depicted in
Fig.~\ref{fig:geometry}.  The electric field with amplitude $E$ points in
the $x$~direction, and the magnetic field with amplitude $B$ in the
$y$~direction.  This configuration corresponds to the quasistatic limit
of either a plane-wave field or two counterpropagating laser fields.  In
the case of a plane-wave field, its wave vector will be directed along
the positive $z$~direction, and the amplitudes of the electric and
magnetic fields will be equal.  The superposition of two
counterpropagating laser fields can lead to orthogonal electric and
magnetic fields of different magnitudes.  Note that the orthogonality of
the considered setup is maintained under Lorentz boosts because $\vect
E\cdot \vect B=0$ is Lorentz invariant.  Furthermore, because $\vect
B^2- \vect E^2$ is invariant under Lorentz boosts, the relative strength
of the electric and magnetic fields is also maintained.  This will lead
us later to the distinction between the cases $|E|>|B|$, $|E|=|B|$, and
$|E|<|B|$.  Note that for $|E| > |B|$, one can always boost along the
$z$~direction into a new reference frame where $B' = 0$.  Similarly for
$|E| < |B|$, there exists a reference frame where $E' = 0$.  For $|E| =
|B|$, all boosts along the $z$~direction maintain the condition $|E'| =
|B'|$.

The wave vector of a possibly assisting high-energy photon is parallel
to the $z$~direction; i.\,e., it may be positive or negative.  Other
relative orientations of the photon and the electromagnetic field are
not considered here because such a setup would lead to a reduced or even
vanishing pair production rate.  

In the following sections, we will utilize a semiclassical description
of pair creation based on classical trajectories.  For an economical
description of these trajectories, the following notation will be
used:  The electron's kinetic four-momentum is written as
\begin{equation}
  p^\mu = (p_0, p_x, p_y, p_z)\,, 
\end{equation}
and likewise its canonical momentum $P^\mu$.  The kinetic and canonical 
momentum of the positron are denoted by $q^\mu$ and $Q^\mu$, respectively.  The
wave vector for the photon $k^\mu$ is written, according to the geometry used
in this work, as
\begin{equation}
  k^\mu = (k_0, 0, 0, k_z) 
\end{equation}
with $k_0=|k_z|$.  As the motion of the particles can be reduced to a
one-dimensional description along the $x$~direction, we denote the
electron's and positron's $x$~coordinates by $x^-$ and $x^+$ (not to be
confused with light-cone coordinates).  Variables and their values at the
point of pair production $x_s$ will be subindexed by ``s'', thus
$p^\mu_s = p^\mu(x_s)$ or $x^-_s$ being the $x$~component of the
electron's position at $x_s$.  Variables and their values at the point
where the electron or the positron leave the imaginary trajectory will
be subindexed by ``e'' (for exit), thus $p_{x,e} = p_x(x^-_e)$ being the
kinetic momentum of the electron in the $x$~direction at its point of
exit.

The equations of motion for the canonical momenta of the electron and the
positron with charge $\mp e$ are given by 
\begin{subequations}
  \label{eqn:eoms}
  \begin{align}
    \frac{\dd}{\dd t} P^{\mu}  & = 
    -e\, \frac{\partial x^{\nu}}{\partial t}
    \frac{\partial A_{\nu}}{\partial x_{\mu}} \,,\\
    \frac{\dd}{\dd t} Q^{\mu}  & = 
    e\, \frac{\partial x^{\nu}}{\partial t}
    \frac{\partial A_{\nu}}{\partial x_{\mu}} \,,
  \end{align}
\end{subequations}
where $A_{\nu}$ denotes the electromagnetic field's four-potential.  The
electron's and the positron's kinetic and canonical momenta are connected via
\begin{subequations}
  \label{eqn:canon_kin_rel}
  \begin{align}
    p^{\mu}(x) & = P^{\mu}(x) + e A^{\mu}(x) \,,\\
    q^{\mu}(x) & = Q^{\mu}(x) - e A^{\mu}(x) \,.
  \end{align}
\end{subequations}

\subsection{Matrix elements for pair creation}

The mathematical handle for pair production is given by its so-called
matrix element or transition amplitude from an initial vacuum state
(possibly including a photon) to a final state with an electron and
positron (see Ref.~\cite{gitman1991quantum} for a thorough
treatment).  For an external electromagnetic field only, this is given by
\begin{subequations}
  \begin{equation}
    M_\text{fi} = \scalprod{1_p, 1_q, \text{out}}{0, \text{in}} \,.
    \label{eqn:mfi_wo_photon}
  \end{equation}
  Both the initial in-state and the final out-state are defined in the
  Furry picture \cite{furry1951onbound} and refer to a common time (see
  also Appendix~\ref{appx:qft_to_classic} or
  Ref.~\cite{gitman1991quantum}).  The asymptotic four-momenta of the
  electron and positron are indicated by $p$ and $q$.

  In the case of an additional quantized photon field, which may assist
  the process, the matrix element will read in first-order perturbation
  theory
  \begin{equation}
    M_\text{fi} = i \int \dd x \bra{1_p, 1_q, \text{out}} \op{\mathcal{H}}_\text{int}(x)
    \ket{1_k, \text{in}} \,.
    \label{eqn:mfi_w_photon}
  \end{equation}
\end{subequations}
The four-momentum of the quantized photon is labeled by $k$, and
$\op{\mathcal{H}}_\text{int}$ designates the QED interaction vertex.
Using semiclassical methods, the exponential parts of both matrix
elements can be evaluated approximately.  Here, semiclassical refers to
the fact that only classical trajectories (although possibly imaginary)
connecting the in- and the out-states are taken into account in the
path-integral picture.  As shown in Appendix~\ref{appx:qft_to_classic}
(see Eqs.~(\ref{eqn:omega_final}) and (\ref{eqn:mfi_sc_photon})), this
approximation yields
\begin{equation}
  M_\text{fi} \sim \exp \left[ - \Im (W_k + W_p + W_q) \right] \,.
  \label{eqn:mfi_exp_phase_fixme}
\end{equation}
$W_p$ and $W_q$ are the gauge-dependent classical actions of the
electron and positron with asymptotic momentum $p$ and $q$ in the
external field.  Likewise, $W_k$ gives the classical action of the
quantized photon with momentum $k$.  Note that both transitions (with or
without an additional high-energy photon) may coexist.  The limit $k
\rightarrow 0$ leads to a vanishing amplitude in
Eq.~(\ref{eqn:mfi_w_photon}) due to prefactors, but the exponential
approaches the same value as the exponential in
Eq.~(\ref{eqn:mfi_wo_photon}).  Therefore, in the case of no additional
photon, $W_k$ in Eq.~(\ref{eqn:mfi_exp_phase_fixme}) is set to zero.  The
various actions in \eqref{eqn:mfi_exp_phase_fixme} are not the same as
the commonly used actions $S(x',x)$, which connect a position eigenstate
with another position eigenstate.  Rather, they are Legendre transforms
thereof, connecting a position eigenstate with a momentum eigenstate
($\vect{x}'$ being implicitly defined by the canonical momenta, see
Appendix~\ref{appx:qft_to_classic}):
\begin{subequations}
  \begin{align}
    W_p(x) & =  S_p(x', x) - \vect{P} \cdot\vect{x}'\,, \\
    W_q(x) & =  S_q(x', x) - \vect{Q} \cdot\vect{x}'\,, \\
    W_k(x) & =  S_k(x, x') + \vect{K} \cdot\vect{x}'\,.
  \end{align}
\end{subequations}

The classical trajectories will consist of a path of the incoming
photon toward the point of pair creation $x_s$, where the photon
converts into an electron and positron, and two outgoing paths of the
created particles from $x_s$ onward.  If there is no photon, the pair is
created out of the vacuum at $x_s$.  The exponent of
Eq.~(\ref{eqn:mfi_exp_phase_fixme}) is therefore the imaginary part of the
action, acquired by a photon coming from the past and propagating with momentum
$k$ to $x_s$, and the two actions, acquired by the electron and positron
propagating from $x_s$ to the future with momenta $p$ and $q$.  Although the
classical actions are gauge dependent, the square modulus of
Eq.~(\ref{eqn:mfi_exp_phase_fixme}) is not, as the boundary terms at $x_s$
cancel each other and the boundary terms at $\pm \infty$ result in unimportant
phases.  Due to the Lorentz invariance of the actions,
Eq.~(\ref{eqn:mfi_exp_phase_fixme}) is also invariant under Lorentz
transformations.

\subsection{Kinetic considerations at the point of pair production}

At the point of pair production $x^\mu_s$, the classical energy-momentum
conservation
\begin{equation}
  p^\mu_s + q^\mu_s = k^\mu
  \label{eqn:xs_cond}
\end{equation}
must be satisfied for the trajectory being classical.  This cannot happen
on real classical trajectories, but on imaginary ones.  Squaring both
sides of $k^\mu - p^\mu_s = q^\mu_s$ leads to
\begin{equation}
  \label{eqn:kps_equal_zero}
  k p_s = 0 = k_0 p_{0,s} - k_z p_{z,s} \,,
\end{equation}
yielding with $k_0=|k_z|$
\begin{equation}
  \label{eqn:p0_pz}
  p_{0,s}^2 = p_{z,s}^2 \,,
\end{equation}
and likewise for $q^\mu_s$.  Hence, using the relativistic dispersion
relation and squaring $p_s$ and $q_s$ gives
\begin{equation}
  m^2 = -p_{x,s}^2 - p_{y,s}^2 = -q_{x,s}^2 - q_{y,s}^2 \,.
\end{equation}
Due to the fact that $p_y$ and $q_y$ are constants of motion in the
aforementioned setup, they have to be real to be consistent with a real
asymptotic momentum.  Therefore, $p_{x,s}$ and $q_{x,s}$ must be purely
imaginary:
\begin{align}
  \label{eqn:pxs_imag}
  p_{x,s} & = -q_{x,s} = \pm i \sqrt{\smash[b]{m^2 + p_y^2}} = \pm i m_* \,,\\
  m_* & \equiv \sqrt{\smash[b]{ m^2 + p_y^2}} \,. 
\end{align}
Starting from $x_s$, both the trajectory of the electron and the
trajectory of the positron need to be followed until the exit points in
order to calculate the imaginary part of the action along these
trajectories.  The trajectories will be called ``imaginary'', as long as
the momenta $p_x$ and $q_x$ are still imaginary.  Hence, the exit points
are given by the condition that $p_x$ and $q_x$ become zero, and
consequently, real:
\begin{equation}
  p_{x,e} = 0 = q_{x,e} \,.
  \label{eqn:exit_cond}
\end{equation}

\section{Tunneling picture for the constant field approximation}
\label{sec:tunneling_picture}

The tunneling picture, presented in this work, is based on the
assumption that the external electromagnetic field can be treated as
constant and uniform during the imaginary part of the trajectory.  In
Sec.~\ref{sec:Kinetics} this assumption will be discussed in more
detail, and the kinetic equations necessary for determining the pair
creation probability will be derived.  Based on these findings, an
intuitive picture of pair creation will be derived in
Sec.~\ref{sec:Graphical_interpretation} and discussed in detail in the
remaining subsections.  In order to establish this enhanced picture it
is necessary to study the semiclassical trajectories during pair
creation in detail.

\subsection{Kinetics within the constant field approximation}
\label{sec:Kinetics}

Expanding the electromagnetic field in the region of interest just up to
linear terms in $x^\mu$ leads to a constant and uniform electric and
magnetic field.  For the case of an external field due to two
counterpropagating lasers, the approximation is justified if the
spacetime region of the imaginary dynamics, which is $|x^\mu_s -
x^\mu_e|$, is small compared to the spacetime region on which the
electromagnetic field varies.  In the case of an external plane-wave
field, this approximation is also valid if the imaginary trajectory is
such that both the electron and the positron move closely to the plane
wave's phase and therefore see the same field everywhere.  Hence,
applying the constant field approximation requires us to check, after
calculating the trajectories, that along these, the electromagnetic
field under consideration can really be treated as constant.  In
general, the electromagnetic field is evaluated at complex spacetimes.
Using the constant field approximation, the vector potential for the
electromagnetic field may be written as
\begin{equation}
  \label{eqn:vect_pot_arb}
  A_\mu(t,z) = (0, E t - B z, 0, 0) \,.
\end{equation}
Here, $E$ and $B$ correspond to the values of the electric and magnetic
fields at the point where the electromagnetic field is expanded.  For
crossed or counterpropagating laser beams, it might be possible that
$|E| \ne |B|$.  Thus allowing for arbitrary values of $E$ and $B$, there
are three different cases: $\norm{E} > \norm{B}$, $\norm{E} = \norm{B}$,
and $\norm{E} < \norm{B}$.  The corresponding Lorentz-invariant quantity
\begin{equation}
  \mathcal{E} \equiv \sqrt{\norm{E^2 - B^2}} 
\end{equation}
will also be used in this work. \pagebreak[2]

Integrating the equations of motion (\ref{eqn:eoms}) with constant
crossed fields of the geometry in Fig.~\ref{fig:geometry} yields
\begin{subequations}
  \label{eqn:eoms_int_z0}
  \begin{align}
    \label{eqn:eoms_int_p0}
    p_0(x^-) & = p_{0,e} - e E (x^- - x^-_e) \,, \\
    \label{eqn:eoms_int_pz}
    p_z(x^-) & = p_{z,e} - e B (x^- - x^-_e) \,, \\
    \label{eqn:eoms_int_q0}
    q_0(x^+) & = q_{0,e} + e E (x^+ - x^+_e) \,, \\
    \label{eqn:eoms_int_qz}
    q_z(x^+) & = q_{z,e} + e B (x^+ - x^+_e) \,.
  \end{align}
\end{subequations}
Thus, both the kinetic energy and the $z$~momentum depend only linearly
on the position in the $x$~direction, simplifying the calculation
considerably.  In contrast to Schwinger pair creation with an electric
field only, the electron and the positron will be accelerated along the
$z$~direction in the presence of a magnetic field, see
Eqs.~(\ref{eqn:eoms_int_pz}) and (\ref{eqn:eoms_int_qz}).  Furthermore,
the magnetic field guides both particles in the same direction, while the
electric field accelerates the particles into opposite directions.  The
$x$~dependence for $p_x$ and $q_x$ is given implicitly by the dispersion
relation $p^2 = m^2$:
\begin{subequations}
  \label{eqn:eoms_int_x}
  \begin{align}
    p_x^2(x^-) & = p_0^2(x^-) - \left(m^2_* + p_z^2(x^-)\right) \,, \\
    q_x^2(x^+) & = q_0^2(x^+) - \left(m^2_* + q_z^2(x^+)\right) \,.
  \end{align}
\end{subequations}
Using Eq.~(\ref{eqn:eoms_int_z0}) and the relation from
Eq.~(\ref{eqn:kps_equal_zero}), the length of the imaginary trajectory
from $x_s$ to $x_e$ along the $x$~direction for both the electron and
the positron is determined by
\begin{subequations}
  \label{eqn:len_im_traj_x}
  \begin{align}
    x^-_s - x^-_e & = 
    \frac{k_0 p_{0,e} - k_z p_{z,e}}{e (E k_0 - B k_z)} \,, \\
    x^+_e - x^+_s & = 
    \frac{k_0 q_{0,e} - k_z q_{z,e}}{e (E k_0 - B k_z)} \,.
  \end{align}
\end{subequations}
Plugging Eq.~(\ref{eqn:eoms_int_z0}) into Eq.~(\ref{eqn:xs_cond}) gives
the relations
\begin{subequations}
  \label{eqn:eoms_rel1}
  \begin{align}
    \label{eqn:eoms_rel1a}
    e E (x^+_e - x^-_e) & = p_{0,e} + q_{0,e} - k_0 \,, \\
    \label{eqn:eoms_rel1b}
    e B (x^+_e - x^-_e) & = p_{z,e} + q_{z,e} - k_z \,.
  \end{align}
\end{subequations}
Multiplying (\ref{eqn:eoms_rel1a}) with $B$, (\ref{eqn:eoms_rel1b}) with
$E$, and subtracting both yields
\begin{equation}
  B\left( p_{0,e} + q_{0,e} - k_0 \right) = 
  E\left( p_{z,e} + q_{z,e} - k_z \right) \,.
  \label{eqn:pqk_rel}
\end{equation} 
This equation gives the relation between $p^\mu_e$, $q^\mu_e$, and
$k^\mu$, as they are not independent of each other.

As a consequence of Eqs.~(\ref{eqn:eoms_int_z0}) and
(\ref{eqn:eoms_int_x}), the kinetic momenta of the particles depend only
on the $x$~direction of space.  Furthermore, it turns out that the
imaginary part of the exponent is solely determined by the momenta in
the $x$~direction \footnote{This can be shown best in the gauge $A^\mu =
  (-E x, 0, 0, -B x)$ for the general constant crossed field case.} via
\begin{equation}
  W = 
  \int_{x^-_e}^{x_s} p_x \,\dd x + \int_{x^+_e}^{x_s} q_x \,\dd x =
  W^- + W^+ \,,
  \label{eqn:exp_pxdx}
\end{equation}
although the particles' (imaginary) dynamics can be three-dimensional.
The imaginary part is made up from a part due to the electron moving
between $x_s$ and $x^-_e$, where $p_x$ is imaginary, and a part due to
the positron moving between $x_s$ and $x^+_e$, where $q_x$ is
imaginary.  The photon's action cancels with the electron's and the
positron's actions along the $z$~and $t$~directions due to energy and
momentum conservation.  The photon will, however, influence the whole
dynamics, and in this way the momenta $p_x$ and $q_x$.  The integrals for
the exponent Eq.~(\ref{eqn:exp_pxdx}) can be calculated analytically
using Eq.~(\ref{eqn:eoms_int_x}) for the momenta $p_x$ and $q_x$ and
(\ref{eqn:len_im_traj_x}) for the integration limits.  Note that $W^+$
and $W^-$ are not independent of each other due to momentum conservation
at the point of pair creation, see Eq.~\eqref{eqn:xs_cond}.

\subsection{Graphical interpretation of the relativistic tunneling picture}
\label{sec:Graphical_interpretation}

\begin{figure}
  \centering
  \includegraphics[width=0.9\linewidth]{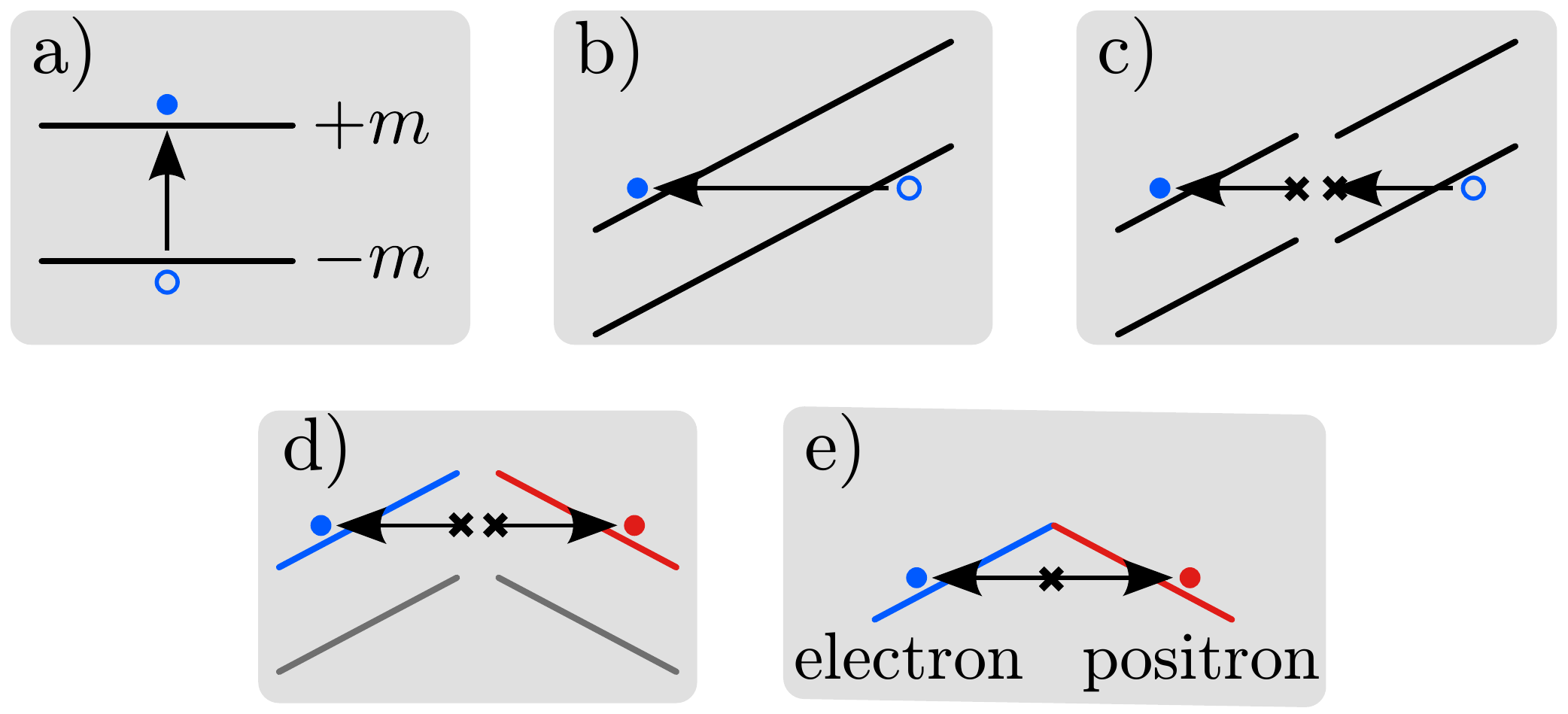}
  \caption{(color online) The conventional tunneling picture of pair
    creation as tunneling from the Dirac sea (a) without electric field
    and (b) with electric field.  The tunneling path may be split into
    two parts (c), which may be interpreted as an electron and a
    positron emerging in the barrier and tunneling to real
    positive-energy states (d and~e).}
  \label{fig:tp_o2n}
\end{figure}

Pair creation in strong electric fields is commonly interpreted as
tunneling from a negative-energy state in the Dirac sea to a
positive-energy state \cite{popov1972pair}.  In other words, the
creation of an electron-positron pair is described by a transition from
a one-particle state to another particle state.  This picture can,
however, be translated into a picture where tunneling starts under the
tunneling barrier and the final state after tunneling is a classical
two-particle state, see Fig.~\ref{fig:tp_o2n}.  Although both
interpretations are equivalent, a pair creation picture that involves
two particles may be more intuitive.  In the following, a pair creation
picture will be introduced that goes beyond the standard tunneling
picture of Ref.~\cite{popov1972pair} by incorporating also an external
magnetic field and possibly a high-energy photon.  This picture is
inspired by the tunneling picture for atomic ionization.

The standard tunneling picture for atomic ionization gives an intuitive
measure for the suppression of ionization by means of an area that is
determined by the potential function and the particle's energy.  If the
potential is increased, for example, this area gets larger, thus
indicating a reduced tunneling probability.  In the same way, the
enhanced tunneling picture of pair creation presented in this work will
give a measure for the suppression of pair production by means of an
area.  The exact exponents are given by the integrals in
Eq.~(\ref{eqn:exp_pxdx}), which can be pictured by areas given by the
integrals of
\begin{subequations}
  \label{eqn:im_px_qx}
  \begin{align}
    \Im p_x(x) & = \sqrt{\smash[b]{m^2_* + p_z^2(x) - p_0^2(x)}} \,, \\
    \Im q_x(x) & = \sqrt{\smash[b]{m^2_* + q_z^2(x) - q_0^2(x)}}
  \end{align}
\end{subequations}
along the $x$~direction.  Plotting these quantities directly would yield
the exact phase-space areas of the tunneling trajectories, but
unfortunately, it does not give any intuitive account on how the pair
production changes if, for example, the electric or magnetic field
amplitude or the energy of the additional photon changes.  However, it
will be sufficient to use quantities that are monotonically correlated
to the momenta in Eq.~(\ref{eqn:im_px_qx}).  These quantities will be
called $\tilde{p}_x$ and $\tilde{q}_x$ and will increase or decrease if
$p_x$ and $q_x$ increase or decrease.  This monotonic correlation assures
that the areas spanned by $\tilde{p}_x$ and $\tilde{q}_x$ along the
$x$~direction also increase or decrease if the exponents in
Eq.~(\ref{eqn:exp_pxdx}) increase or decrease.  Although the choice for
$\tilde{p}_x$ and $\tilde{q}_x$ is not unique, we defined them as
\begin{subequations}
  \label{eqn:approx_pxqx_measure}
  \begin{align}
    \Im{} \tilde{p}_x & = 
    \left| 
      \sqrt{\smash[b]{m^2 + p_x^2 + p_y^2 + p_z^2}} - 
      \sqrt{\smash[b]{m^2 + p_y^2 + p_z^2}} 
    \right| \nonumber \\
    & = \norm{p_0 - \tilde{p}_0} = \tilde{p}_0 - p_0 \,, \\
    \Im{} \tilde{q}_x & = 
    \left| 
      \sqrt{\smash[b]{m^2 + q_x^2 + q_y^2 + q_z^2}} - 
      \sqrt{\smash[b]{m^2 + q_y^2 + q_z^2}} \right| \nonumber \\
    & = \norm{q_0 - \tilde{q}_0} = \tilde{q}_0 - q_0 \,,
  \end{align}
\end{subequations}
with the pseudoenergies $\tilde{p}_0$ and $\tilde{q}_0$ defined as
\begin{subequations}
  \label{eqn:pseudo_kin_energy}
  \begin{align}
    \tilde{p}_0(x) & = \sqrt{\smash[b]{m^2 + p_y^2 + p_z^2(x)}} = 
    \sqrt{m^2_* + p_z^2(x)} \,, \\
    \tilde{q}_0(x) & = \sqrt{\smash[b]{m^2 + q_y^2 + q_z^2(x)}} = 
    \sqrt{m^2_* + q_z^2(x)} \,.
  \end{align}
\end{subequations}
The areas spanned by $\tilde{p}_x$ and $\tilde{q}_x$ along the
$x$~direction are the areas enclosed by the curves of the kinetic energy
and the pseudoenergy, both amenable for an intuitive interpretation.
The monotonic correlation between the imaginary part of $\tilde{p}_x$
and $p_x$ can be shown by taking the derivative of $\tilde{p}_x$ with
respect to $p_x$, which is always positive.  This holds similarly for
the monotonic correlation between the imaginary part of $\tilde{q}_x$
and $q_x$.  Furthermore, $\tilde{p}_x$ and $\tilde{q}_x$ are zero if
$p_x$ and $q_x$ are zero, which corresponds to the exit points, and
hence the exit points are automatically given by the points of
intersection between the kinetic and pseudoenergy curves.  Despite the
fact that this approximation to the area of the exponents is not exact,
it is sufficient to derive qualitative results and gives more intuition,
as we will show in the following.

\begin{figure}
  \centering
  \includegraphics[width=0.925\linewidth]{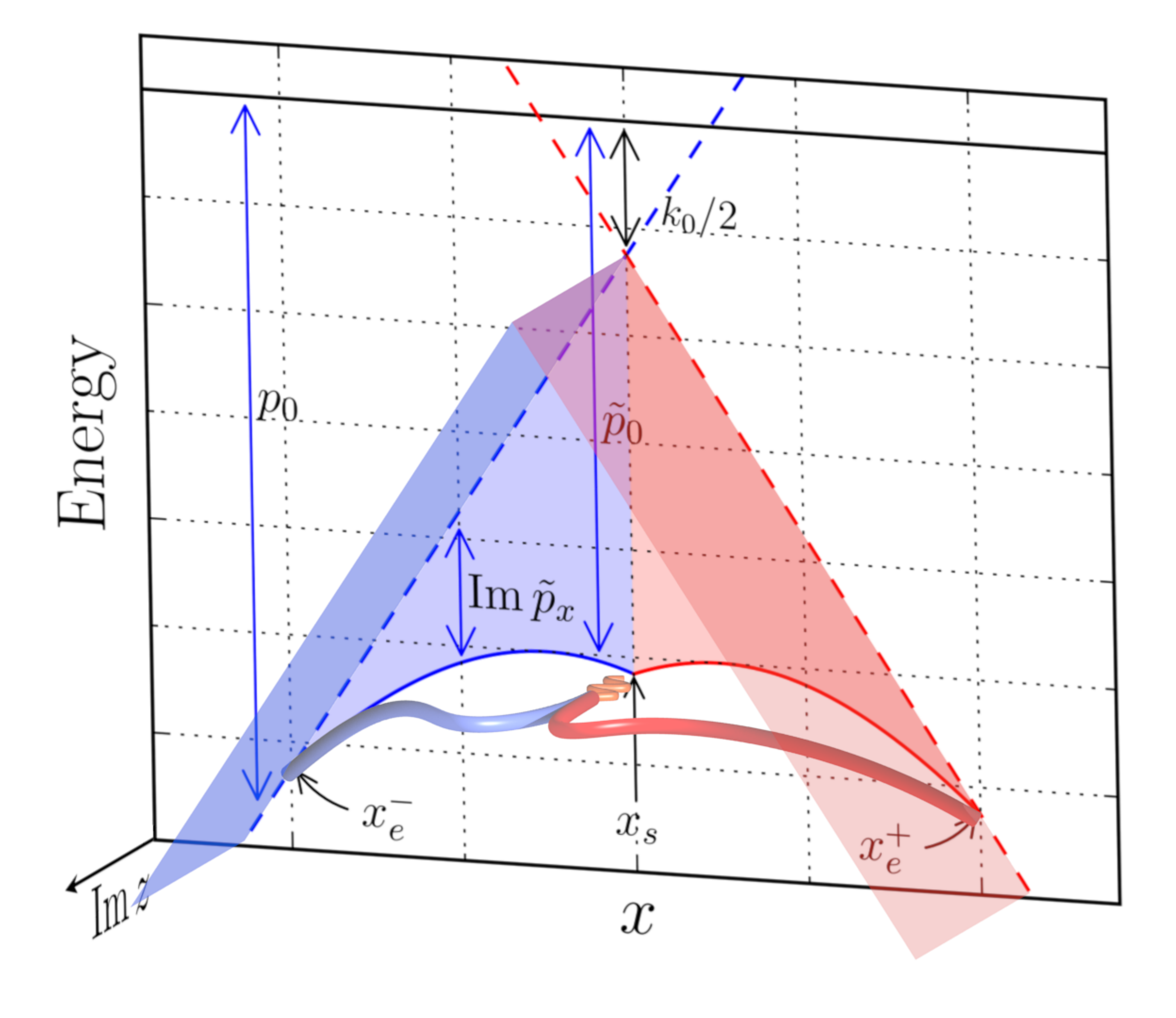}
  \caption{(color online) The tunneling picture of pair creation,
    indicating also the quantities used in the text.  The electron and
    the positron start their imaginary trajectory at $x_s$ and travel to
    their respective exits $x^-_e$ and $x^+_e$, where the trajectories
    become real.  The tunneling dynamics is along the real $x$~axis and
    the presence of a magnetic field may cause a nontrivial motion along
    the imaginary $z$~axis.  The bold solid lines represent the
    pseudoenergies in Eq.~(\ref{eqn:pseudo_kin_energy}) as a function of
    the $x$ and $z$~coordinates, whereas the thin solid lines indicate
    the projection of Eq.~(\ref{eqn:pseudo_kin_energy}) along the
    $x$~axis.  The dashed lines represent the kinetic energy.  The
    values of the acquired imaginary actions $W^-$ for the electron and
    $W^+$ for the positron along these paths determine the exponential
    term suppressing pair production and can be inferred qualitatively
    by the shaded areas, enclosed by the thin solid and dashed red and
    blue lines.}
  \label{fig:3d}
\end{figure}

\begin{figure*}
  \centering
  \includegraphics[width=0.925\linewidth]{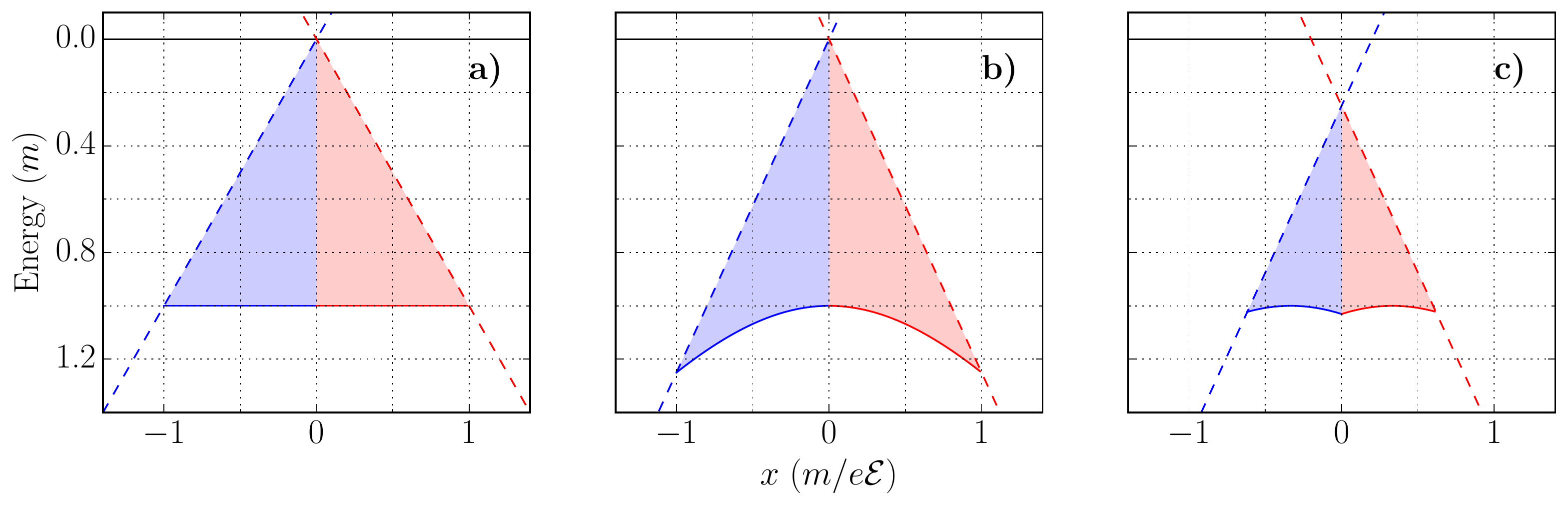}
  \caption{(color online) The tunneling picture of pair creation for
    $|E|>|B|$.  (a) The standard Schwinger case with an electric field
    only is considered.  %
    (b) An inertial system, which is Lorentz-boosted relatively to that
    of (a) along the $z$~direction with $v/c = -3/5$ leading to a
    nonzero magnetic field.  The larger (boosted) electric field leads to
    a steeper increase of the kinetic energy (dashed line).
    Furthermore, the magnetic field increases the pseudoenergy (solid
    lines) due to the buildup of momentum along the $z$~direction.
    (c) The same configuration as in part (b), but with an additional
    photon of momentum $k_0 = m/2$.  Its kinetic energy is shared by the
    created electron and positron, shifting their kinetic energies
    (dashed lines) down by $k_0/2$ and in this way decreasing the shaded
    area.  Also, the kinetic momentum of the photon is shared by the
    produced electron and the positron shifting the initial pseudoenergy
    (solid lines) at $x_S$ downwards.}
  \label{fig:case1_expl}
\end{figure*}

The now-introduced quantities can be presented graphically as a function
of the $x$~coordinate giving a visual representation of the tunneling
picture of pair creation, as shown exemplarily in Fig.~\ref{fig:3d}.
The upper black solid line represents the zero-energy reference.  Lines
for the electron are drawn in blue, whereas lines for the positron are
drawn in red.  The bold solid colored lines represent the particles'
pseudoenergies in Eq.~(\ref{eqn:pseudo_kin_energy}) as a function of the
particles' real $x$~coordinate and the imaginary $z$~coordinate.  The
motion into the imaginary $z$~direction results from a real $z$~momentum
due to the presence of a magnetic field or the initial momentum transfer
due to the additional photon during an imaginary time interval.  Note,
however, that there is no motion in the $z$ direction in real space
because tunneling is instantaneous in real time \cite{klaiber2013under}.
Because the semiclassical tunneling trajectories do not represent
realistic particle paths, the exit coordinate, which is where the
particles' momenta become real, is not necessarily real, i.\,e., $\Im
z_e^\pm\ne0$.  This may be related to the fact that each semiclassical
trajectory corresponds to a delocalized quantum state with a
well-defined momentum, which is given by the exit momentum of the
semiclassical trajectory.  A real motion into the $z$~direction would
require an inhomogeneous electromagnetic field similarly to the
near-threshold-tunneling regime of relativistic tunnel
ionization~\cite{yakaboylu2013relativistic}.

The particles' pseudoenergies in Eq.~(\ref{eqn:pseudo_kin_energy}) as
well as their kinetic energies in Eqs.~\eqref{eqn:eoms_int_p0}
and~\eqref{eqn:eoms_int_q0} can be expressed as functions of the
$x$~coordinate only and are represented in Fig.~\ref{fig:3d} by the
solid and dashed lines, respectively.  The solid colored lines originate
for both the electron and the positron from the initial point for pair
creation $x_s$ and end at their respective exit points $x^-_e$ and
$x^+_e$, where they intersect the dashed lines.  Hence, by following, for
example, the blue dashed and solid lines, the change in kinetic energy
and pseudoenergy can be traced along the $x$~direction of the imaginary
trajectory from the point of pair creation until the exit for the
electron, and similarly for the positron by following the red lines.
Finally, a blue shaded area for the electron and a red shaded area for
the positron are shown, which are enclosed by the dashed and solid
colored lines.  These areas can be related to exponents for pair
production.

After introducing the different constituents of the enhanced tunneling
picture, their physical interpretation can be explained now.  At the
point of pair creation $x_s$, the sum of the kinetic energies of the
electron and the positron must be equal to the initial energy already
existing in the system.  This is either zero or $k_0$ in the case of an
additional photon.  From $x_s$, both the electron and the positron
follow their imaginary trajectories until their respective exits.  The
trajectories will be imaginary as long as $p_x$ and $q_x$ are imaginary
or equivalently as long as the pseudoenergies $\tilde{p}_0$ and
$\tilde{q}_0$ are larger than their respective kinetic energies $p_0$
and $q_0$.  In the graphs this can be seen by the solid colored lines
lying below the dashed colored lines.  At the exits, $p_x = 0$ and $q_x
= 0$, both lines intersect as both types of energy (pseudo and kinetic)
become equal.  Note that the kinetic energies go below their minimum
allowed energy behind the exit points towards $x_s$, where $p_x$ and
$q_x$ get imaginary.  It is just this behavior that allows the
fulfillment of Eq.~(\ref{eqn:xs_cond}), i.e., energy-momentum
conversation at $x_s$.  Furthermore, the difference between the
pseudoenergy and kinetic energy is taken as an approximate measure for
the imaginary part of the particles' $x$~momenta, see
Eq.~(\ref{eqn:approx_pxqx_measure}).  The exponent for pair production,
Eq.~(\ref{eqn:exp_pxdx}), is given by the integral over $p_x$ and $q_x$
along the $x$~direction from the point of pair creation $x_s$ to both
the exit points of the electron and positron.  Hence, the absolute value
of these exponents can be inferred approximately by the shaded areas in
the graphs.  Thus, the shaded blue area corresponds to $W^-$ and the
shaded red area corresponds to $W^+$.  In the following three
subsections, the qualitative behavior of these shaded areas---that is,
the pair production exponent---will be discussed depending on different
factors like the ratio of electric and magnetic field strength or the
impact of an additional photon.

\subsection{\texorpdfstring{Case \boldmath$\norm{E} > \norm{B}$}{E larger than
B}}

Figure~\ref{fig:case1_expl} shows some common cases where the electric
field is stronger in magnitude than the magnetic field.  It can be
readily seen that for these cases, pair production is always possible
due to the existence of the exit points, given by the intersection of
the dashed and solid lines.  For $|E| > |B|$, this intersection is
always possible, because the pseudoenergy line (solid) never falls more
steeply than the kinetic energy line (dashed), and hence, they need to
intersect somewhere.

\begin{figure*}
  \centering
  \includegraphics[width=0.925\linewidth]{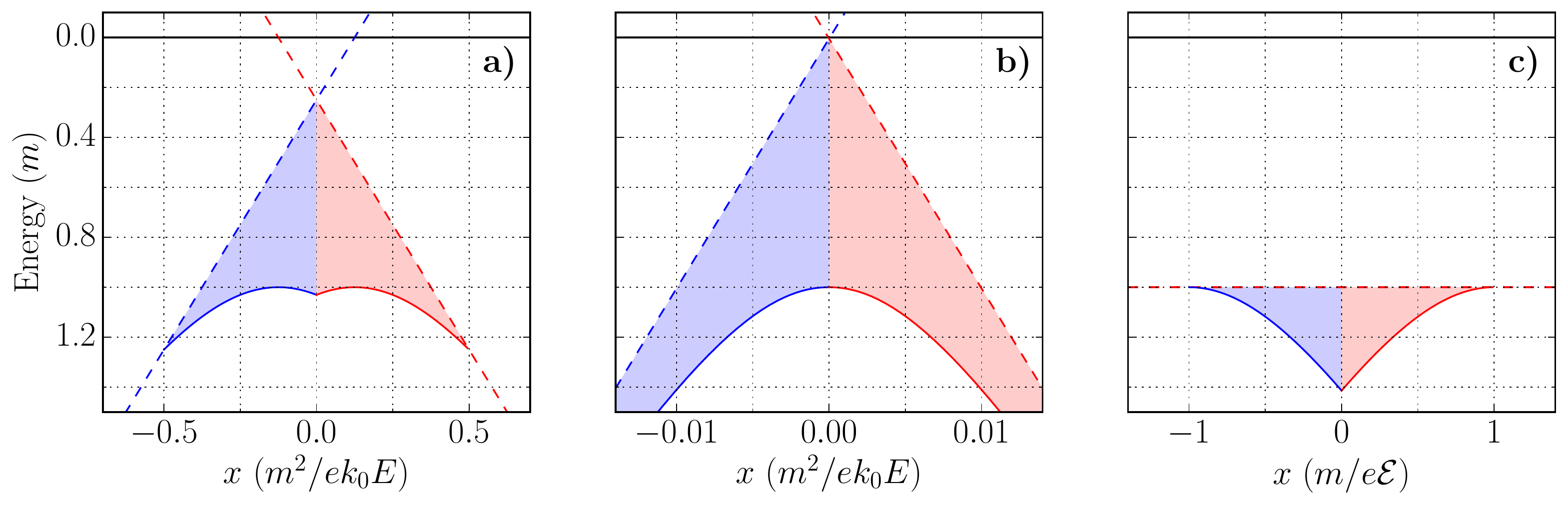}
  \caption{(color online) Part (a) and (b) show the tunneling picture
    for the plane-wave case $|E| = |B|$, with an additional photon with
    $k_0 = m/2$ and $k_0 = m/100$, respectively, similarly to
    Fig.~\ref{fig:case1_expl}(c).  In this case, an additional photon
    propagating opposite to the external electric field is needed for
    pair creation.  Without this photon, the solid and dashed lines
    become tangent asymptotically, yielding an infinite area and hence
    no pair production.  From a kinetic point of view, the magnetic
    field (being as strong as the electric field) builds up as much
    $z$~momentum (increasing the pseudoenergy) as the electric field
    kinetic energy, and hence the particles cannot become real.  The
    energy shift by the photon is enough to let both lines intersect
    each other, provided that its $z$~momentum is opposite to the
    buildup due to the magnetic field.  The case with a magnetic field
    only is shown in part (c).  To render pair production possible, the
    incoming photon must have at least 2 times the rest mass energy,
    i.\,e., $k_0 \ge 2 m$, because the magnetic field only cannot
    transfer energy.  It can be seen from the diagram that the initial
    $z$~momentum, transferred from the photon to the electron and
    positron, is rotated along the imaginary trajectory onto the
    $x$~momentum and hence reduces the initial imaginary $x$~momentum to
    zero at the exits, making the electron and positron real, and
    therefore pair production possible.}
  \label{fig:case23}
\end{figure*}

In Fig.~\ref{fig:case1_expl}(a), Schwinger pair creation is considered
with an electric field only and without an additional photon.
Furthermore, the most probable case with zero momentum at the exit is
taken.  The pseudoenergies $\tilde{p}_0$ and $\tilde{q}_0$ (the solid
colored lines) are constant, because $p_z$ and $q_z$ do not change when
there is no magnetic field, and $p_y$ as well as $q_y$ are constants of
motion.  At $x_s$, both kinetic energies of the electron and the
positron are zero in sum, as there is no additional energy without a
photon.  The Lorentz-boosted version of Fig.~\ref{fig:case1_expl}(a),
seen in a reference frame boosted along the $z$~direction, is shown in
Fig.~\ref{fig:case1_expl}(b).  Due to the boost, the electric field gets
larger, and hence the dashed colored lines get steeper.  The stronger
electric field alone would increase the tunneling probability.  But due
to the boost, a magnetic field is also experienced now by the particles,
which transfers momentum from the $x$~direction to the $z$~direction.
This momentum transfer reduces the acceleration into the electric field
direction and makes pair creation less probable.  In the tunneling
picture, the momentum transfer is represented by the bent pseudoenergies
$\tilde p_0$ and $\tilde q_0$; see Fig.~\ref{fig:case1_expl}(b).
Furthermore, the energy of the created particles is increased, which is
consistent with the boost in the $z$~direction.  The growth of the
relativistic mass is seen by the bending of the pseudoenergy lines.
Both effects, the increase of the electric field and the increase of the
relativistic energy, cancel each other exactly, as the transition
amplitude is Lorentz invariant.  This invariance is also represented in
Figs.~\ref{fig:case1_expl}(a)~and~\ref{fig:case1_expl}(b), since the
size of the shaded areas remains almost constant under the Lorentz
boost, although the areas are a qualitative measure for the tunneling
probability.  Note also that the $x$~coordinate of the exit points
$x^-_e$ and $x^+_e$ does not change, as it is not affected by a boost
along the $z$~direction.  Furthermore, Fig.~\ref{fig:case1_expl}(b)
allows us to infer the effect of a magnetic field that is superimposed
to a given electric field.  The area in Fig.~\ref{fig:case1_expl}(b) is
increased due to the bending of the solid lines opposed to the case
without magnetic field, where these lines are horizontal.  Consequently,
the pair production rate is decreased.

In Fig.~\ref{fig:case1_expl}(c), an additional photon is incorporated
into the setting of $E$ and $B$ in Fig.~\ref{fig:case1_expl}(b).  Here,
two very important things can be seen immediately.  First, the
additional energy of $k_0$ at $x_s$.  Both the electron and the positron
can now share this energy, which is $k_0/2$ per particle for maximum
pair production probability.  For this reason, the dashed colored lines
are shifted down by $k_0/2$.  This decreases the shaded areas and
therefore increases the pair production probability.  Secondly, the
additional photon does not only transfer its kinetic energy; it also
transfers its momentum along the $z$~direction.  This leads to an
additional $z$~momentum of $k_0/2$ for both the electron and the
positron, increasing their pseudoenergy, or equivalently, their
relativistic mass at $x_s$.  Without a magnetic field, this extra energy
due to the $z$~momentum remains until the exit and must be supported by
the electric field, ultimately decreasing the pair production
probability due to a longer ``time'' until the exit.  To summarize, the
additional photon transfers energy, which enhances the probability, but
it also transfers momentum, which degrades the probability.  In total,
this results always in an enhanced probability.  The degradation due to
the momentum transfer along the $z$~direction can be reduced by tuning
the magnetic field in such a way that it decelerates the particles along
the $z$~direction, thus making them ``lighter'' at the exits.  The
appropriate tuning of the magnetic field will be discussed in more
detail in Sec.~\ref{sec:max_prob_traj}.  From Eq.~(\ref{eqn:eoms_rel1})
it can be shown that the direction in which the electron and the
positron are accelerated along the $x$~direction is determined only by
the orientation of the electric field,
\begin{equation}
  \sgn{(x^+_e - x^-_e)} = \sgn{E}\,.
  \label{eqn:case1_x_dir}
\end{equation}
This is represented in the tunneling picture by the slope of the kinetic
energy lines being positive or negative, forcing the particles in the
one or the other direction.  

Calculating the tunneling probability via the integral in
Eq.~(\ref{eqn:exp_pxdx}) gives for the electron
\begin{gather}
  W^- = i \frac{(E p_{0,e} - B p_{z,e})^2}{2
    \mathcal{E}^3} 
  \left( \arccos \Gamma - \Gamma \sqrt{1-\Gamma^2} \right)
  \label{eqn:wminus_e}
\intertext{with}
  \Gamma  = 1 - e \frac{E^2 - B^2}{E p_{0,e} - B p_{z,e}} 
  (x^-_s - x^-_e)\,.
\end{gather}
The quantity $W^+$ has the same analytical structure, but the momenta
and the position of the electron need to be replaced by that of the
positron.  For the standard Schwinger case, which is just an electric
field without an additional photon and with zero momentum at the exit,
$W^-$ in Eq.~(\ref{eqn:wminus_e}) reduces to $i\pi m^2/(4E)$.  Adding
$W^+$ and multiplying by~2, which corresponds to the square modulus of
the matrix element, gives exactly the Schwinger exponent $\pi m^2/E$.
Furthermore, if the electrical field strength $E$ is replaced by its
corresponding Lorentz invariant $\mathcal{E}$, then this result also
agrees with the first-order exponent in the work \cite{nikishov1970pair}
for nonzero magnetic fields in the case of no additional photon.

\subsection{\texorpdfstring{Case \boldmath$\norm{E} = \norm{B}$}{E equals B}}

This case is somewhat special, because it connects the previous and the
next case in a singular manner.  It corresponds to an external
plane-wave field under the constant crossed field approximation.
Figures~\ref{fig:case23}(a) and~\ref{fig:case23}(b) show the tunneling
picture for a quantized photon of energy $k_0 = m/2$ and $k_0 = m/100$.
In general, the picture is quite similar to that of
Fig.~\ref{fig:case1_expl}(c), but here a quantized photon is mandatory
for pair production.  Due to the magnetic field being as strong as the
electric field, the energy transferred to the particles in $z$~momentum
by the magnetic field is as strong as the kinetic energy transferred to
the particles by the electric field.  In this sense, the electric field
without an additional photon is not strong enough to let the particles
leave the imaginary trajectories, as the magnetic field makes them more
and more heavy.  This can be seen by comparing in
Figs.~\ref{fig:case23}(a) and~\ref{fig:case23}(b) how the colored dashed
and solid lines approach each other.  In the limit of $k$ going to zero,
the kinetic and pseudoenergy lines become tangent and will not
intersect, leading to an infinite exponent and thus no pair production.

Depending on the orientation of $E$ and $B$, which is connected with the
orientation of $k_\text{L}$ for a plane wave ($\vect{E}\times \vect{B}$
points into the same direction as $\vect{k}_L$), the quantized photon
has to be directed antiparallel to the external field wave vector.  In
the given geometry, this means that
\begin{equation}
  \sgn{k_z} = - \sgn{(E B)}
\end{equation}
has to be satisfied.  Otherwise, there will be no pair production, as
the quantized photon and external field will be parallel, which can be
thought of as one plane-wave field, which is known to produce no pairs.
If the additional photon propagates into the opposite direction, then
the pseudoenergy lines falls down such that it do not cross the kinetic
energy lines, in contrast to the case in Fig.~\ref{fig:case23}(a).  This
is because the photon momentum points into the direction in which the
magnetic field accelerates the particles, leading to no intersection and
hence infinite suppression of pair creation.  As in the previous case,
$|E|>|B|$, the direction of movement for the electron and positron along
$x$ is given by Eq.~(\ref{eqn:case1_x_dir}).

The imaginary part of the exponent for the electron now reads
\begin{equation}
  W^- = i\frac{2}{3} \frac{
  \left[2 e(E p_{0,e} - B p_{z,e}) (x^-_s - x^-_e) \right]^{3/2}}{
  \norm{2 e(E p_{0,e} - B p_{z,e})}} \,.
  \label{eqn:exp_case2}
\end{equation}
For comparison, the same result will be derived in
Appendix~\ref{appx:pert_eval} by a different approach.  Instead of the
constant field approximation, the action integral $W$ will be evaluated
perturbatively in terms of the classical nonlinearity parameter
$\xi=eE/(m\omega_L) \gg 1$ for the special case of an external
plane-wave field with the characteristic frequency $\omega_L$.  For
maximum probability, the total exponent for pair production reduces to
the value $4 m^3/(3 e k_0 E)$.  This is in accordance with the value
already calculated in Ref.~\cite{reiss1962absorption}.

\subsection{\texorpdfstring{Case \boldmath$\norm{E} < \norm{B}$}{E less than B}}

For this case, there is a minimal energy the quantized photon must have
in order to produce a pair.  Boosting into a reference frame, where the
electric field vanishes, it is clear that no work will be done by the
electromagnetic field.  Hence, the whole energy in this frame for the
pair must be carried by the quantized photon.  In the case of no
electric field and both the electron and positron carrying only their
rest mass after creation, the quantized photon has to carry an energy of
exactly $2 m$, as shown in Fig.~\ref{fig:case23}(c).  Both kinetic
energy lines are horizontal because of the electric field being zero, and
their sum is equal to the initial photon energy of $2 m$.  If this
initial energy would be smaller than $2 m$, then the pseudoenergy lines
would not hit the kinetic energy lines; instead, they would stay below
them.  This minimum energy condition is also given by
Eq.~(\ref{eqn:pqk_rel}), which yields for $E = 0$
\begin{equation}
  k_0 = p_{0,e} + q_{0,e} \,.
\end{equation}
The effect of a nonzero electric field may also be deduced from
Fig.~\ref{fig:case23}(c).  A nonzero electric field tilts the kinetic
energy lines to the one or the other direction, and consequently the exit
points will be closer to or farther from $x_s$, and hence pair production
will be enhanced or suppressed.

Interpreting Fig.~\ref{fig:case23}(c) in terms of imaginary
trajectories, it can be understood in the following way.  At the point
of pair production $x_s$, the momentum along the $x$~direction for the
electron and the positron has the value $\pm i m$.  The momentum along
the $z$~direction, transferred by the photon, is $m$.  Following the
imaginary trajectories from $x_s$ on, the magnetic field rotates the
momentum along the $z$~direction onto the $x$~direction and cancels
exactly the initial imaginary momentum.  Thus, the particles become real
and pair production is possible.  If the initial energy of the photon is
smaller, then the magnetic field cannot turn the particles into real
ones.  If there is also an electric field, which reduces $p_x$ and $q_x$
in the same direction as the magnetic field, then the particles can
leave the imaginary trajectory sooner, making pair production more
probable.  Pair production will be reduced or even completely cut off if
the electric field works in the opposite direction, and therefore, the
energy carried by the photon does not suffice anymore.

\begin{figure*}
  \centering
  \includegraphics[width=0.925\linewidth]{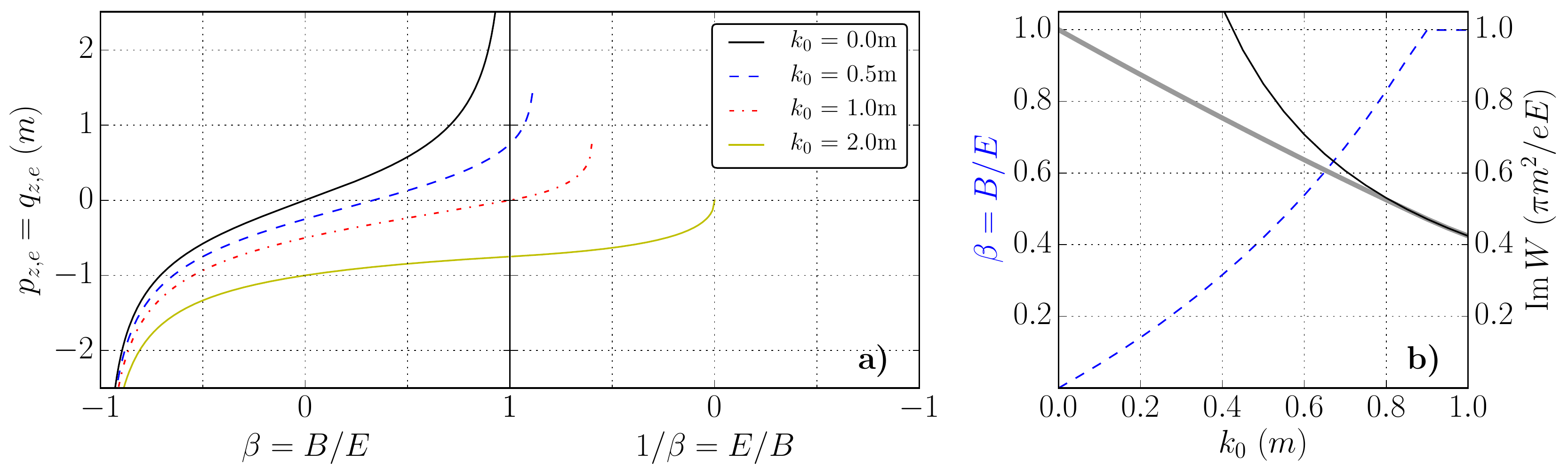}
  \caption{(color online) Properties of the maximum probability
    trajectories.  (a)~The most probable $z$~momentum at the tunneling
    exit of the electron and positron depending on the ratio $\beta =
    E/B$ and photon energy $k_0$.  The $y$~momentum increases the
    relativistic mass of the particles and is therefore zero for maximum
    probability.  By definition, the $x$~momentum must be zero at the
    exit.  For $\beta = 0$ (no magnetic field), the most probable
    $z$~momentum is exactly half the photon momentum $k_0$.  Changing
    the magnetic field will introduce a shift of the momentum depending
    on the sign of $\beta$.  In the right pane of part (a), where $|B| >
    |E|$, the lines end at some $0<1/\beta<0$.  This is due to the
    energy cutoff depending on the energy of the photon.
    (b)~The optimum ratio $\beta$ for a given photon energy $k_0$ to
    achieve maximal pair production probability.  The dashed blue line
    gives the optimum ratio depending on $k_0$, while the bold gray line
    gives its corresponding value for the exponent.  At $k_0 = 0$, for
    example, it is best to have no magnetic field (Schwinger case).  At
    $k_0 = k_0^* \equiv \sqrt{\smash[b]{4/5}}\, m$, the ratio $\beta$
    becomes~1, corresponding to the plane-wave case.  The exponent for
    the plane-wave case is also given for reference by the thin black
    line.  For $k_0 < k_0^*$, a tuned magnetic field can therefore
    enhance the pair production probability, while for values greater
    than $k_0 = k_0^*$, $\beta = 1$ is always optimal for a given fixed
    maximum field strength.}
  \label{fig:max_plts}
\end{figure*}

In contrast to the previous two cases, the direction of tunneling is now
determined by the orientation of the $B$ field and the quantized photon
\begin{equation}
  \sgn{(x^+_e - x^-_e)} = -\sgn{(B k_z)} \,.
\end{equation}
If the electric field works in the tunneling direction, then the
probability will be enhanced; otherwise it will be reduced.  For
$|E|<|B|$, the exponent for the electron computes to
\begin{equation}
  W^- = 
  i \frac{(E p_{0,e} - B p_{z,e})^2}{2\mathcal{E}^3} 
  \left( \arcosh \Gamma - \Gamma \sqrt{\Gamma^2 - 1} \right)\,.
  \label{eqn:exp_case3}
\end{equation}

\section{Maximum probability trajectories}
\label{sec:max_prob_traj}

Each tunneling trajectory is characterized by its final momenta of the
created electron and positron and has a specific pair production
probability, which depends on an exponential term with exponent $\Im
W=\Im W^++\Im W^-$ given in Eqs.~(\ref{eqn:wminus_e}),
(\ref{eqn:exp_case2}) and~(\ref{eqn:exp_case3}) as well as on a
prefactor.  The maximum probability trajectories are defined as the
trajectories, where the exponential term $\Im W$ is minimized for given
electromagnetic field strengths $E$ and $B$ and a given photon energy
$k_0$.  In general, the most probable trajectories are always the most
symmetric ones.  For the kinetic momenta at the exit, this means
\begin{subequations}
  \begin{align}
    \label{eq:pex_sym}
    p_{x,e} & = q_{x,e} = 0 \,, \\
    \label{eq:pey_sym}
    p_{y,e} & = q_{y,e} = 0 \,, \\
    \label{eq:pez_sym}
    p_{z,e} & = q_{z,e} \,,
  \end{align}
\end{subequations}
where (\ref{eq:pex_sym}) follows from the definition of the exit point
and (\ref{eq:pey_sym}) is required by minimizing the kinetic energy of
the created particles.  

The kinetic momenta $p_{z,e}$ and $q_{z,e}$ depend on the ratio $\beta =
B/E$ as well on the momentum of the photon $k_0$.
Figure~\ref{fig:max_plts}(a) illustrates the $\beta$-dependence of
$p_{z,e}$ and $q_{z,e}$ for various $k_0$.  In the left part of
Fig.~\ref{fig:max_plts}(a) $\norm{E} > \norm{B}$, while in the right
part $\norm{E} < \norm{B}$.  At the center and at the outer borders, $E$
and $B$ have the same magnitude.  For the Schwinger case with $B = 0$
and $k_0 = 0$, the final momentum of the maximum probability
trajectories is zero, as shown in the figure.  Varying $\beta$ while
keeping $k_0 = 0$ yields nonzero momenta $p_{z,e}$ and $q_{z,e}$, see
solid black line in Fig.~\ref{fig:max_plts}(a).  The momentum $p_{z,e}$
corresponds exactly to the $\beta$ boosted momentum of the zero momentum
of $B = 0$.  In the limit $|\beta|\to 1$, the momenta $p_{z,e}$ and
$q_{z,e}$ and consequently the relativistic masses of the created
particles diverge, which is related to the fact that pair production is
not allowed in this regime without a photon.

In the presence of a photon with momentum, the created electron and
positron share the photon's momentum at $x_s$ if $\beta = 0$ and keep it
until the exit, because there is no magnetic field, and therefore,
$p_{z,e}=q_{z,e}=k_0/2$ as also shown in Fig.~\ref{fig:max_plts}(a).
Going to the left of $\beta = 0$ shows that $p_{z,e}$ and $q_{z,e}$
diverge also for $k_0>0$, because the electric and the magnetic fields
work into different directions.  Going to the right of $\beta = 0$, they
will work into the same region, and now pair production is also possible
in the region $\norm{\beta}>1$, i.\,e., where $\norm{B} > \norm{E}$.
For $k_0 < 2m$, the lines for the momenta $p_{z,e}$ and $q_{z,e}$ end
before $1/\beta = 0$ because the photon's energy is not enough to create
pairs without an electric field.  The line for $k_0 = 2m$ ends exactly
at $1/\beta = 0$ with $p_{z,e} = 0$.  This corresponds to the case
discussed in Fig.~\ref{fig:case23}(c).  Going further to the right of
$1/\beta = 0$, the electric and magnetic field will work into opposite
directions, and in this way render pair production impossible.

For a given photon momentum $k_0$, pair creation may be maximized by
varying $\beta$ and minimizing the imaginary part of the exponent for
the maximum probable trajectories.  The corresponding exponent is shown
in Fig.~\ref{fig:max_plts}(b).  The blue dashed line indicates the optimal
ratio $\beta$.  At $k_0 = 0$, it is optimal to have only an electric
field.  For $k_0>0$ also, the optimal $\beta$ is nonzero, meaning that a
magnetic field will enhance pair production.  If the photon momentum is
larger than the critical momentum
\begin{equation}
  k_0^* \equiv \sqrt{\smash[b]{4/5}}\,m \approx 0.89\,m
  \label{eqn:k_star}
\end{equation}
then the optimal $\beta$ is~1, which corresponds to a plane wave.  This
means that, for a fixed maximum field strength, the plane-wave field is
always optimum for photon energies larger than $k_0^*$, in the setup
treated in this work.  The value (\ref{eqn:k_star}) may be calculated by
taking the derivative of Eq.~(\ref{eqn:wminus_e}) with respect to the
magnetic field.  This derivative is evaluated for the limit $B
\rightarrow E$ and then set to zero, which gives an implicit equation
for $k_0^*$ that can be solved analytically.  Note that $k_0^*$ does not
depend on the electromagnetic field magnitude $E$.  The imaginary values
of the exponent $W$ are also shown in Fig.~\ref{fig:max_plts}(b) for the
plane-wave case (black solid line) and for the optimal case (gray solid
line).  The exponent for the optimum tuned case starts at $k_0 = 0$ with
$i\pi m^2/(eE)$ (Schwinger case) and then decreases with increasing
$k_0$ until it coincides with the value of the exponent for the
plane-wave case at $k_0 = k_0^*$.

\section{Conclusion} 
\label{sec:conclusions}

We introduced an intuitive tunneling picture for pair creation, which
also incorporates effects due to a magnetic field and an additional
high-energy photon.  This picture is Lorentz invariant and does not
depend on a particular gauge and is valid for homogeneous constant
electromagnetic fields.  This constant field approximation is valid,
e.\,g., for the long wave limit of pair creation in counterpropagating
laser fields.

The relativistic tunneling picture can be drawn due to the
quasi-one-dimensional nature of the used setup in the quasistatic limit.
Various features of pair creation can be inferred qualitatively from the
introduced tunneling picture.  For example, an additional photon lowers
the potential barrier due to the photon's energy but also increases the
particle's relativistic mass due to the photon's additional momentum.
Due to this increased relativistic mass, the electron and positron stay
longer (in terms of imaginary time) under the barrier until they gain
enough energy to become real.  An additional magnetic field, however,
will also change the momentum under the barrier, and in this way, the
relativistic mass.  Depending on the magnetic field's direction and
magnitude it may counteract the increase of the relativistic mass due to
the photon's momentum and, therefore, increase the pair production
probability.

The relativistic tunneling picture has been devised on the basis of a
semiclassical approximation using classical trajectories.  This approach
also allowed us to calculate the exponents of the transition amplitudes.
The calculated exponents for maximum probability agree with the
analytical results by Nikishov \cite{nikishov1970pair} in the case of no
incoming photon for the three different possibilitie:s $|E| > |B|$, $|E|
= |B|$, and $|E| < |B|$.  Furthermore, they agree with the results by
Reiss \cite{reiss1962absorption}, assuming a plane-wave external laser
field with an incoming photon in the limit $\xi \ll 1$.

Other geometries, as well as time and space varying fields, may also be
treated in the tunneling regime by imaginary trajectories, but their
respective picture will be inherently multidimensional and hard to
visualize.  Nevertheless, simple trends due to time and space variation
may be interpreted with the current picture, which will be subject to
future work.

\begin{acknowledgments}
  We are grateful to K.~Z.~Hatsagortsyan, A.~Di Piazza, S.~Meuren, and
  E.~Yakaboylu for valuable discussions.
\end{acknowledgments}

\appendix

\section{From quantum field theory to classical actions}
\label{appx:qft_to_classic}

This paragraph recapitulates how to derive the semiclassical approximation for
matrix elements, which is usually symbolically written as
\begin{equation}
  M_{fi} \sim \exp \left( - \Im S \right)\, .
\end{equation}
This form in general is quite ambiguous, because the definition of the
action $S$ itself due to its gauge dependence is not unique, and
furthermore it also depends on the type of initial and final states,
which can be eigenstates of position, momentum or of a specific
Hamiltonian (to name the most common).

Following the treatment in Ref.~\cite{gitman1991quantum}, the mode
Hamiltonian $\mathcal{H}_\text{e}$ of the matter part (e.\,g., electrons
and positrons), describing the free modes of these particles is given by
\begin{equation}
  \mathcal{H}_\text{e} = 
  \vect{\alpha} \left( -i \vect{\nabla} \right) + \beta m \,.
  \label{eqn:free_dirac_ham_s}
\end{equation}
The quantization leads to the following field Hamiltonian for the free
quantized matter field:
\begin{align}
  \op{H}_\text{e} & = \int \dd \vect{x} \, \op{\psi}^+(\vect{x}) \, 
  \mathcal{H}_\text{e} \, \op{\psi}(\vect{x}) \nonumber \\
  & = \int \dd \vect{x} \, \op{\overline{\psi}}(\vect{x}) \,
  \left( -i \vect{\gamma} \vect{\nabla} + m \right) \, \op{\psi}(\vect{x}) \,.
  \label{eqn:free_dirac_ham}
\end{align}
The same procedure applied to the free quantized radiation field leads
to its corresponding field Hamiltonian $\op{H}_\gamma$, which explicit
form is not of interest here.  The interaction between both fields is
given by
\begin{equation}
  \op{H}_\text{int,full}
  = e \int \dd \vect{x} \, \op{\overline{\psi}}(\vect{x}) \,
  \gamma_\mu \, \op{\psi} (\vect{x}) \, \op{A}_{\text{full}}^\mu (\vect{x}) \,.
  \label{eqn:quant_qed_int}
\end{equation}
Both $\op{\psi}(\vect{x})$ and $\op{A}_{\text{full}}^\mu(\vect{x})$ are
the field operators of the quantized matter and radiation field in the
Schr{\"o}dinger picture, whence they only depend on $\vect{x}$.  This theory
describes the full QED.  But, so far, it does not allow to solve even
very simple problems, and hence some approximations need to be done.
The most fundamental approximation, as is common for problems of QED in
external fields, is the separation of the full electromagnetic field
into a quantized radiation field $\op{A}^\mu(\vect{x})$ and a classical
external field $A^\mu_\text{ext}(x)$.  Treating the external field as a
classical field is motivated by the fact that it should be very intense
and quantum effects are negligible.  Its interaction with the matter
field is given by the following Hamiltonian:
\begin{equation}
  \op{H}_\text{int,ext} =
  e \int \dd \vect{x} \, \op{\overline{\psi}}(\vect{x}) \,
  \gamma_\mu \, \op{\psi} (\vect{x}) \, A^\mu_\text{ext} (x) \,.
\end{equation}
Note the explicit time dependence of this Hamiltonian due to the
possible time dependence of $A^\mu_\text{ext}(x)$.  A further important
fact of the external field approximation is neglecting the influence of
the matter field on the dynamics of the external field, also called
backreaction.  That way, $A^\mu_\text{ext}(x)$ is assumed to behave like
a classical free electromagnetic field that does not see the current
\begin{equation}
  \op{j}^\mu = e \, \op{\overline{\psi}} \,
    \gamma_\mu \, \op{\psi}
\end{equation}
created by the matter field.  Putting together $\op{H}_\text{e}$ and
$\op{H}_\text{int,ext}$ gives the (possible time-dependent) quantized
field Hamiltonian for the matter field in an external field,
\begin{align}
  \op{H}_\text{e,ext} & = \op{H}_\text{e} + \op{H}_\text{int,ext} \\
  & = \int \dd \vect{x} \, \op{\overline{\psi}}(\vect{x}) \,
    \Bigl[ \vect{\gamma}\left( -i \vect{\nabla} -
    \vect{A}_\text{ext} \right) + m + e \gamma^0 A_\text{ext}^0 \Bigr] \, \op{\psi}(\vect{x}) 
  \,.
  \nonumber
\end{align}
Thus, the modes of the matter field in the external electromagnetic field
are described by the following Hamiltonian, known as the Dirac
Hamiltonian with an external field:
\begin{equation}
  \mathcal{H}_\text{e,ext} = \vect{\alpha}
  (-i \vect{\nabla} - \vect{A}_\text{ext}) + \beta m + e A^0_\text{ext}\,,
  \label{eqn:mod_ham_dirac_ext}
\end{equation}
and therefore they take the influence of the external field into full
account.  The interaction with the quantized field is still treated
perturbatively.  For doing this, it is convenient to switch to the
interaction picture, with the free part
\begin{equation}
  \op{H}_0 = \op{H}_\gamma + \op{H}_\text{e,ext}
\end{equation}
and the interaction part 
\begin{equation}
  \op{H}_\text{int} = e \int \dd \vect{x} \, \op{\overline{\psi}}(\vect{x}) \,
  \gamma_\mu \, \op{\psi} (\vect{x}) \, \op{A}^\mu (\vect{x}) \,.
  \label{eqn:quant_qed_int_gamma}
\end{equation}
This type of interaction picture, including the interaction with the external
field in $\op{H}_0$, is also called the Furry picture \cite{furry1951onbound}.
The field operator for the matter field becomes time dependent ($\op{U}_0$
being the unitary operator, transforming from the Schr{\"o}dinger to the
interaction picture),
\begin{align}
  \op{\psi}_\text{I}(x) & = \op{U}^+_0 \, \op{\psi}(\vect{x}) \,
  \op{U}^{\phantom{+}}_0 \,,
\end{align}
and it can be shown that this field operator will solve the Dirac equation in
the external field
\begin{equation}
  \left( i \partial_t - \mathcal{H}_\text{e,ext} \right) \op{\psi}_\text{I}(x)
  = 0  \,.
\end{equation}
Knowing this, the field operator $\op{\psi}_\text{I}$ can be expanded
into a complete set of eigensolutions (modes).  There are, of course,
infinitely many different sets.  The two sets ${}_\pm \varphi_n(x)$ and
${}^\pm \varphi_n (x)$ are chosen, satisfying the relations
\begin{subequations}
  \begin{align}
    \mathcal{H}_\text{e,ext}(t_\text{in}) \, {}_\pm
    \varphi_n(\vect{x}t_\text{in})
    & = {}_\pm \varepsilon_n \, {}_\pm \varphi_n(\vect{x}t_\text{in})\,, \\
    \mathcal{H}_\text{e,ext}(t_\text{out}) \, {}^\pm
    \varphi_n(\vect{x}t_\text{out})
    & = {}^\pm \varepsilon_n \, {}^\pm \varphi_n(\vect{x}t_\text{out}) \,.
  \end{align}
\end{subequations}
Thus, the set ${}_\pm \varphi_n(x)$ corresponds to positive/negative-energy
eigensolutions (${}_\pm \varepsilon_n \gtrless 0$) at $t_\text{in}$, and ${}^\pm
\varphi_n$ to positive/negative-energy eigensolutions (${}^\pm \varepsilon_n
\gtrless 0$) at $t_\text{out}$.  Using these sets, the field operator can be
expanded as
\begin{subequations}
  \begin{align}
    \label{eqn:psi_op_dec_in}
    \op{\psi}_\text{I}(x) & = \sum_n \op{a}_n(\text{in}) \, {}_+\!\varphi_n(x)
    + \op{b}^+_n (\text{in}) \, {}_-\!\varphi_n(x) \\
    \label{eqn:psi_op_dec_out}
    & = \sum_n \op{a}_n(\text{out}) \, {}^+\!\varphi_n(x)
    + \op{b}^+_n (\text{out}) \, {}^-\!\varphi_n(x) \,, 
  \end{align}
\end{subequations}
defining also two possible different sets of creation/annihilation
operators, $\op{a}^+$/$\op{a}$ and $\op{b}^+$/$\op{b}$, for positive and
negative energy particles at $t_\text{in}$ and $t_\text{out}$.  These in
turn define the vacuum states at $t_\text{in}$ and $t_\text{out}$:
\begin{subequations}
  \begin{align}
    \op{a}_n(\text{in}) \ket{0\text{, in}} & = 0\,, &
    \op{b}_n(\text{in}) \ket{0 \text{, in}} & = 0 \,, \\
    \op{a}_n(\text{out}) \ket{0\text{, out}} & = 0 \,, &
    \op{b}_n(\text{out}) \ket{0 \text{, out}} & = 0 \,.
  \end{align}
\end{subequations}

\subsection{Pair production due to the external electromagnetic field only}

By considering the external electromagnetic field only (zeroth order in
the quantized radiation field), the time evolution operator in the
interaction picture ($\op{H}_\text{int,I}$ being the interaction picture
representation of Eq.~(\ref{eqn:quant_qed_int_gamma})) reduces to the
identity, that is,
\begin{align}
  \label{eqn:time_op_ip}
  \op{U}(t_\text{out}, t_\text{in}) & = \mathcal{T} \exp \left[ 
  - i \int_{t_\text{in}}^{t_\text{out}} \op{H}_\text{int,I} \, \dd t \right] \\
  & \rightarrow \op{1} \quad \text{(in 0th order)} \,.\nonumber
\end{align}
Thus, as shown in Ref.~\cite{gitman1991quantum}, for all the transition
elements of interest, one only needs to know the following propagator
elements:
\begin{align}
  G(^\zeta|_\varkappa)_{mn} & = 
  \int \dd \vect{x} \,\dd \vect{y} \,
  {}^\zeta\! \varphi_m^+(\vect{x} t_\text{out})
  \, G(\vect{x} t_\text{out}, \vect{y} t_\text{in}) \, {}_\varkappa \! \varphi_n
  (\vect{y} t_\text{in}) \,, \\[1ex]
  G(_\varkappa|^\zeta) & = G(^\zeta|_\varkappa)^+ \,,
\end{align}
with $m$, $n$ being the quantum numbers of the modes at
$t_\text{out/in}$ and $\zeta, \varkappa = \pm$ denoting positive or
negative energy (i.\,e.,~particle or antiparticle).  $G(\vect{x}
t_\text{out}, \vect{y} t_\text{in})$ is the full propagator of
Eq.~(\ref{eqn:mod_ham_dirac_ext}).  These propagator elements represent
the probability that a positive/negative mode from $t_\text{in}$ evolves
into a positive/negative mode at $t_\text{out}$.  According
Ref.~\cite{gitman1991quantum}, the vacuum stability up to zeroth order
in the quantized radiation field is given by the transition amplitude
between $\ket{0 \text{, in}}$ and $\ket{0 \text{, out}}$:
\begin{equation}
  c_\text{v} = \scalprod{0 \text{, out}}{0 \text{, in}} =
  \det G(^+|_+) = \det G(^-|_-) \,.
\end{equation}
Furthermore, the transition amplitude for producing one pair that is
evolving from an initial vacuum state $\ket{0 \text{, in}}$ to a pair state
$\ket{n, m \text{, out}}$ (electron/positron having quantum number $n$/$m$) 
yields
\begin{align}
  \scalprod{n, m \text{, out}}{0 \text{, in}} & = 
  \bra{0 \text{, out}} \! a_m(\text{out}) b_n(\text{out}) \!
  \ket{0 \text{, in}} \nonumber \\
  & \equiv \omega(\overset{+}m \overset{-}n|0) \, c_\text{v} \,,
\end{align}
with
\begin{equation}
  \label{eqn:omega_pm}
  \omega(\overset{+}m \overset{-}n|0) = 
  \left[ G^{-1}(_+|^+) G(_+|^-) \right]_{mn} 
  = - \left[ G(^+|_-) G^{-1}(^-|_-) \right]_{mn} \,. 
\end{equation}

For the semiclassical case, where pair production is exponentially
suppressed, it is possible to write
\begin{subequations}
  \begin{align}
    c_v & = 1 + \mathcal{O}(e^-) \,, \\
    G(^+|_+)_{mn} & = e^{i \theta^+_n} \delta_{mn} +
    \mathcal{O}(e^-) \,, \\
    G(^-|_-)_{mn} & = e^{i \theta^-_n} \delta_{mn} +
    \mathcal{O}(e^-) \,, \\
    G(^+|_-)_{mn} & = \mathcal{O}(e^-) \,, \\
    G(^-|_+)_{mn} & = \mathcal{O}(e^-) \,.
  \end{align}
\end{subequations}
The notation $\mathcal{O}(e^-)$ stands for the exponentially small
corrections.  Physically, this means that the positive/negative modes at
$t_\text{in}$ mainly evolve into their corresponding positive/negative
modes at $t_\text{out}$ up to a phase $\theta^+$/$\theta^-$.  Only an
exponentially small fraction of their norm goes into different modes,
regardless of whether they are positive or negative.  Inverting
$G(^+|_+)$ and $G(^-|_-)$ yields, therefore,
\begin{align}
  G(^+|_+)^{-1}_{mn} & = e^{-i \theta^+_n} \delta_{mn} + 
    \mathcal{O}(e^-) \,, \\
  G(^-|_-)^{-1}_{mn} & = e^{-i \theta^-_n} \delta_{mn} +
    \mathcal{O}(e^-) \,.
\end{align}
Plugging this into the first line of Eq.~(\ref{eqn:omega_pm}) results in
\begin{equation}
  \omega(\overset{+}m \overset{-}n|0) = 
  e^{i \theta^+_m} G(_+|^-)_{mn} + \mathcal{O}^2(e^-) \,.
\end{equation}
Thus, up to first order in the exponential suppression, the single pair
production elements in Eq.~(\ref{eqn:omega_pm}) are given by the matrix
$G(_+|^-)$ or alternatively by $G(^+|_-)$.  The matrix element $G(_+|^-)_{mn}$
gives the overlap between the negative mode ${}^-\varphi_n$,
propagated from $t_\text{out}$ to $t_\text{in}$, with the positive
mode ${}_+\varphi_m$ at $t_\text{in}$.  Explicitly, this reads
\begin{equation}
  G(_+|^-)_{mn} =  \int \! \dd \vect{x}' \,\dd \vect{x} \, 
  {}_+\!\varphi_m^+(\vect{x}') \,
  G(\vect{x}' t_\text{in}, \vect{x} t_\text{out}) \,
  {}^-\!\varphi_n(\vect{x}) \,.
\end{equation}%
\begin{figure}
  \centering
  \includegraphics[width=0.8\linewidth]{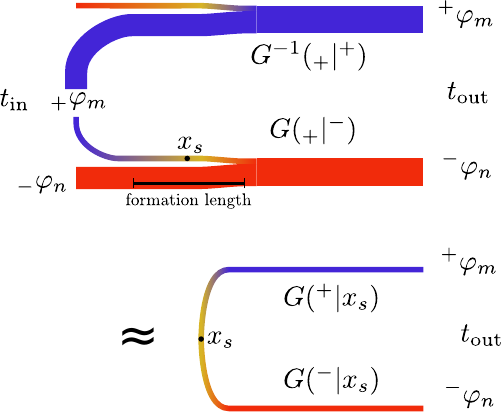}
  \caption{Visualization of the single pair production element of
    Eq.~(\ref{eqn:omega_pm}).  Both positive and negative modes (bold)
    propagate back in time.  Only an exponentially suppressed part
    (thin) splits up and evolves into different modes.  The different
    modes under consideration must match at $t_\text{in}$.  In the
    semiclassical approximation, the transition element can be
    approximated by the classical trajectory, that connects both the
    electron and positron at $t_\text{out}$ with their initial point of
    creation $x_s$.  The formation length gives the typical timescale
    that is needed for splitting up a different mode from the main mode}
  \label{fig:qft2ca}
\end{figure}%
This means that somewhere in between $t_\text{out}$ and $t_\text{in}$, a
small part of ${}^-\varphi_n$ splits up and evolves into
${}_+\varphi_m$.  Still, the main part goes into ${}_-\varphi_n$
($G(_-|^-)$).  This is shown schematically in the upper part of
Fig.~\ref{fig:qft2ca}.  Also shown there is a length, the so-called
formation length, because typically the timescale on which the
conversion from ${}_-\varphi_n$ to ${}^+\varphi_m$ happens is finite.
By finite, we mean that only exponentially small corrections occur if
the formation length is increased further.  In this sense, the formation
length is not strictly defined; rather, it gives some estimate for the
timescale on which most of the conversion has already happened (see also
Ref.~\cite{nikishov1985problems} for the case of a constant
electromagnetic field).

The real space propagator is approximated in the semiclassical
approximation by
\begin{equation}
  G(\vect{x}' t', \vect{x} t) \sim
  \exp \left[ i S(\vect{x}' t', \vect{x} t ) \right] \,.
\end{equation}
Similarly, the propagator from a momentum eigenstate to another (in our
notation taking the quantum numbers $n = \vect{p}$ and $m = \vect{q}$) is
approximated semiclassically as
\begin{equation}
  G\left( \vect{p} t', \vect{q} t \right) \sim
  \exp \left[ i S(\vect{x}' t', \vect{x} t) - i \vect{P} \vect{x}' 
    + i \vect{Q} \vect{x} \right]
  \label{eqn:qc_mom_prop}
\end{equation}
with $\vect{x}'$ and $\vect{x}$ implicitly given by
\begin{equation}
  \vect{P} = \left. \frac{\partial S}{ \partial \vect{x}'} \right|_{\vect{x}'}
  \quad \text{and} \quad 
  \vect{Q} = \left. \frac{\partial S}{ \partial \vect{x}} \right|_{\vect{x}} .
\end{equation}
The additional terms are due to the in- and out-modes being momentum and
not position eigenstates, see also
Refs.~\cite{popov1968quasiclassical,marinov1972pair}.  Thus, the full
propagator gets approximated only by its classical trajectory connecting
the appropriate in- and out-modes instead of all trajectories in the
path integral picture.  One peculiarity of Eq.~(\ref{eqn:qc_mom_prop})
is the matrix structure of its left-hand side corresponding to the full
propagator.  Its right-hand side does not have this matrix structure, as
the exponential is a scalar.  As shown in Ref.~\cite{pauli1932diracs},
the semiclassical treatment of the Dirac equation leads to a prefactor
with matrix structure, which must not be necessarily diagonal in the
zeroth order of $\hbar$.  Therefore, a transition from a positive-energy
spinor to a negative-energy spinor is possible.  Nevertheless, the main
contribution to the matrix elements stems from the exponential factor.
In the classical picture, the transition from a negative-energy
eigenstate to a positive must happen on an imaginary trajectory, because
an electron with positive energy will always stay an electron with
positive energy on a real trajectory.  But if it goes over to an
imaginary path and wraps around the root of its kinetic energy $p_0 =
\sqrt{m^2 + \vect{p}^2}$, it is possible that it continues on the other
branch of the square root---that is, $p_0 = -\sqrt{m^2 +
  \vect{p}^2}$---and therefore becomes a negative energy electron.  This
point on the imaginary part of the trajectory, where this transition
occurs, might be defined by $p_0 = 0$ and will be denoted by $x_s$.  It
should be mentioned that it has no direct physical meaning; i.\,e., it
cannot be measured.  Nevertheless, it can be related to the region of
the formation length.  Using this point as the point of pair production
in the classical picture, the propagator in Eq.~(\ref{eqn:qc_mom_prop})
can be split into
\begin{equation}
  G(\vect{p}t_\text{in}, \vect{q} t_\text{out}) \sim
  e^{i S(\vect{x}'t_\text{in}, x_s) - i \vect{P} \cdot\vect{x}'} 
  e^{i S(x_s, \vect{x} t_\text{out}) + i \vect{Q} \cdot\vect{x}}\,,
  \label{eqn:rs_prop_split}
\end{equation}
where the back part corresponds to a negative energy electron traveling
back in time from $t_\text{out}$ to $x_{0,s}$, which can also be
interpreted as a positron going from $x_{0,s}$ to $t_\text{out}$
(changing sign in $\vect{q}$ and charge).  For the full matrix element
in Eq.~(\ref{eqn:omega_pm}), the front part of
Eq.~(\ref{eqn:rs_prop_split}) gets multiplied by $G(^+|_+)^{-1}$.  In
our approximation this corresponds to a propagation of the
positive-energy electron back from $t_\text{in}$ to $t_\text{out}$, and
one can therefore write
\begin{equation}
  G(^+|_+)^{-1} e^{iS(\vect{x}' t_\text{in}, x_s) - i \vect{P} \cdot\vect{x}'}
  \sim e^{i S(\vect{x}' t_\text{out}, x_s) - i \vect{Q} \cdot\vect{x}'} \,.
\end{equation}
Accordingly, the transition element in Eq.~(\ref{eqn:omega_pm}) can now be
written as
\begin{equation}
  \omega(\overset{+}m \overset{-}n|0) = 
  \exp\left[ - \Im (W_p + W_q) \right] \,,
  \label{eqn:omega_final}
\end{equation}
where $W_p = S(\vect{x}' t_\text{out}, x_s) - \vect{P} \cdot\vect{x}'$ and
$W_q = S(\vect{x} t_\text{out}, x_s) - \vect{Q} \cdot\vect{x}$ correspond
to the modified actions of the electron and positron along their
classical imaginary trajectories.  This result is schematically drawn in
the lower part of Fig.~\ref{fig:qft2ca} with
\begin{math}
  G({}^+|x_s) \sim e^{i W_p} 
\end{math}
and
\begin{math}
  G({}^-|x_s) \sim e^{i W_q} 
\end{math}.
If there are, for a specific field configuration, different imaginary paths
possible for connecting the electron and positron in their specific out-states,
their contribution must be added and might lead to interference phenomena.

\subsection{Pair production due to perturbative treatment of the quantized electromagnetic field}

For this case, the first-order correction to the time evolution operator
in Eq.~(\ref{eqn:time_op_ip}) is given by
\begin{align}
  \op{U}\left( t_\text{out}, t_\text{in} \right) & \rightarrow
  -i \int_{t_\text{in}}^{t_\text{out}} \op{H}_\text{int,I} \,\dd t \nonumber \\
  & \rightarrow -i e \int \op{\overline{\psi}}_\text{I}(x) \, \gamma^\mu \,
  \op{\psi}_\text{I}(x) \op{A}^\mu_\text{I}(x) \, \dd x\,.
\end{align}
The field operators in their interaction representation can be expanded (as
before) into the solutions of their corresponding mode Hamiltonians.  For the
matter field operator $\op{\psi}_\text{I}$, the decomposition
Eq.~(\ref{eqn:psi_op_dec_out}) for the out-states is used again, as the main
interest lies in pair creation and not pair annihilation.  The field operator
for the free quantized electromagnetic field is decomposed as
\begin{equation}
  \label{eqn:emf_op_dec}
  \op{A}^\mu(x) = \sum^3_{\lambda = 0} \int \left[ 
  \op{c}^{\phantom{+}}_{\vect{k} \lambda} f^\mu_{\vect{k}\lambda}(x) +
  \op{c}^+_{\vect{k} \lambda} f^{\mu*}_{\vect{k} \lambda} (x) \right] \dd
  \vect{k}\,,
\end{equation}
and $f^\mu_{\vect{k}\lambda}(x)$ being the photon wave function
proportional to \footnote{As we do not consider spin effects in this
  paper, the given actions do not include any spin or polarization
  terms.}
\begin{equation}
  f^\mu_{\vect{k}\lambda}(x) \sim \exp[-i k x] = \exp\left[ i W_k(x) \right],
\end{equation}
where $W_k(x) = - kx$ usually is called its classical action.  Care should be
taken here, since this action does not correspond to an action of the type
$S(x', x)$ but rather to the modified action discussed before, where on one
side the momentum $k$ and on the other side the spacetime $x$ is fixed.

Expressing also the solutions ${}^\pm\varphi_n$ to the Dirac equation in
their semiclassical approximation and setting the final quantum number
$n$ to momentum eigenstates ($p$ for the electron and $q$ for the
positron), they read ${}^+\varphi_n(x) \sim \exp[ -i W_p(x)]$ and
${}^-\varphi_n(x) \sim \exp [ + i W_q(x)]$ for the electron and the
positron, respectively.  The ``+'' in the positron's exponent is due to
its quantization as an antiparticle.  Again, $W_p$ and $W_q$ are usually
called the electron and positron classical actions, but the same
distinction as mentioned before applies here.  Note that in the case of
an external plane-wave field, these actions are given by the so-called
Volkov actions, and the semiclassical exponent is exact.

By taking into account the first order of the interaction with the quantized
field, pair production can now happen with the additional help of an incoming
quantized photon.  The in-state is defined as 
\begin{equation}
  \ket{0, 0, 1_k} \equiv \op{c}^+_{\vect{k}} \ket{0 \text{, in}} \,,
\end{equation}
and the out-state as
\begin{equation}
  \ket{1_p, 1_q, 0} \equiv 
  \op{a}^+_p(\text{out}) \op{b}^+_q(\text{out}) \ket{0 \text{, out}} \,.
\end{equation}
The transition amplitude for this process to first order is given by
\begin{equation}
  M_{fi} = -ie \bra{1_p, 1_q, 0} \! \int \op{\overline{\psi}}_\text{I}(x) \,
  \gamma^\mu \, \op{\psi}_\text{I}(x) \op{A}^\mu_\text{I}(x) \, \dd x
  \ket{0, 0, 1_k}.
\end{equation}
Expanding the field operators according to
Eqs.~(\ref{eqn:psi_op_dec_out}) and~(\ref{eqn:emf_op_dec}) and writing
(approximating) the field mode solutions in terms of their modified
classical action $W$ yields
\begin{equation}
  M_{fi} \sim - ie \, c_v \int \dd x \, e^{\left[ i \left( 
  W_k(x) + W_p(x) + W_q(x) \right) \right]} \,.
  \label{eqn:mfi_exp_phase}
\end{equation}
Applying the method of stationary phase to this integral yields the
following condition for a stationary phase:
\begin{align}
  0 & = \frac{\partial}{\partial x^\mu} \left(W_k + W_p + W_q\right)
  \!\vert_{x_\text{s}} \nonumber \\[1.5ex]
  & = K^\mu(x_\text{s})
  - P^\mu(x_\text{s}) - Q^\mu(x_\text{s}) \nonumber \\[1.5ex]
  & = k^\mu(x_\text{s}) - p^\mu(x_\text{s}) - q^\mu(x_\text{s}) \,.
  \label{eqn:state_phase_cond}
\end{align}
The last line follows from the previous one due to the fact that the
gauges at the same spacetime point ($x_\text{s}$) cancel each other
exactly \footnote{Using usual quantization rules, the photon will have
  no gauge, and therefore $K = k$.  Furthermore, $P = p - e A$ and $Q =
  q + e A$, which gives $ P + Q - K = p + q -k$.}.  Note that the
modified actions $W$, which depend on momentum and space, are used in
the derivation of Eq.~(\ref{eqn:state_phase_cond}) and not the usual
actions $S$, which depend on two positions.  Although the modified
actions $W$ are gauge dependent, the point of stationary phase $x_s$ is
given by a gauge-independent and Lorentz-invariant kinetic equation,
which can be interpreted from a classical point of view.  According to
Eq.~(\ref{eqn:mfi_exp_phase}), the matrix element is an interference
from all spacetime points $x$ where the photon converts into electron
and positron obeying the asymptotic momenta.  The outcome of this
interference, e.\,g., the matrix element, can be approximated by the
point of stationary phase.  Due to energy-momentum conservation $p + q =
k$ at this point, the laws of classical mechanics are satisfied.
Therefore, the particle conversion stemming from the creation and
annihilation operators in quantum field theory can be brought to a
classical level in the semiclassical approximation.

The classical trajectory will now look like an incoming photon that will
convert at $x_\text{s}$ instantaneously into an electron and positron
while satisfying energy-momentum conservation.  The electron and
positron will travel further in the external field and approach their
asymptotic momenta.

The condition $p + q = k$ cannot be satisfied in real spacetime, as can
be seen by squaring both sides of this equation, but in imaginary
spacetime.  This leads to an imaginary classical trajectory.  Therefore,
approximating Eq.~(\ref{eqn:mfi_exp_phase}) semiclassically yields
\begin{equation}
  M_{fi} \sim - ie \, c_v \, e^{- \Im( W_k + W_p + W_q )} \,.
  \label{eqn:mfi_sc_photon}
\end{equation}

\clubpenalty = 300
\widowpenalty = 300 
\displaywidowpenalty = 300

\section{Perturbative evaluation of the exponent for the plane-wave case}
\label{appx:pert_eval}

For the plane-wave four-potential, we choose the following gauge (with
$\eta = k_L x$ being the phase of the laser, $E$ the maximum field
amplitude and $\omega_L = k_{L,0}$ the characteristic frequency):
\begin{equation}
  A_\mu(\eta) = \left(0, \frac{E}{\omega_L} f(\eta), 0, 0\right)\,,
  \label{eqn:plane_wave_potential}
\end{equation}
yielding for the electron the constants of motion (according to
Eq.~(\ref{eqn:eoms})) $P_x$, $P_y = p_y$, and $P_0 - P_z = p_0 - p_z$.
The last expression can be written in a Lorentz-invariant way,
$\Lambda_p = k_L p$.  Accordingly, the same is true for the positron.
For the quantized photon, this notation can be used too, $\Lambda_k =
k_L k$, yielding also a constant of motion.

To make use of these constants of motion, the transformation 
\begin{equation}
  \left( t, x, y, z \right) \rightarrow
  \left( \eta, x', y', z' \right) = 
  \left( k_L(t - z), x, y, z/k_L \right) 
\end{equation}
into a different coordinate system is applied ($x$, $y$ and $z$ are
spacetime coordinates and not four-vectors in this paragraph).  The
total differential of the spacetime-dependent part of the action $W$ in
this coordinate system
\begin{equation}
  W(x, y, z, t) = W\left( x(x'), y(y'), z(z'), t(z', \eta) \right)
\end{equation}
is given for the electron and analogously for the positron by%
\begin{multline}
  \dd W(x', y', z', \eta) = \\
  - P_x \dd x' - P_y \dd y' + 
  \omega_L(P_0 - P_z)  \dd z'  + P_0/\omega_L \dd \eta \,,
\end{multline}
which can be written in the gauge of
Eq.~(\ref{eqn:plane_wave_potential}) as
\begin{equation}
  \dd W(x', y', z', \eta) = -P_x \dd x' - P_y \dd y' + \Lambda_p \dd z'
  + \frac{p_0}{\omega_L} \dd \eta \,.
  \label{eqn:plane_wave_diff_action}
\end{equation}
The first three terms are the constants of motion and can be integrated:
\begin{equation}
  W(x', y', z', \eta) = -P_x x' - P_y y' + \Lambda_p z' +
  \int^\eta \frac{p_0}{\omega_L} \, \dd \varphi \,.
  \label{eqn:plane_wave_int_action}
\end{equation}
Similarly, the photon's action can be written this way:
\begin{equation}
  W(x', y', z', \eta) = k_x x' + k_y y' - \Lambda_k z' - \int^\eta 
  \frac{k_0}{\omega_L} \, \dd \varphi .
\end{equation}
Allowing only for the classical trajectory---that is, four-momentum
conservation at the point of pair creation---yields the following set of
constraints:
\begin{subequations}
  \label{eqn:plane_wave_conservation}
  \begin{align}
    P_x + Q_x & = p_x + q_x = 0 \,, \\
    P_y + Q_y & = p_y + q_y = 0 \,, \\
    \label{eqn:lambda_p_q_k}
    \Lambda_p + \Lambda_q & = \Lambda_k \,, \\
    \label{eqn:pw_eta_s}
    p_0(\eta_s) + q_0(\eta_s) & =  k_0\,.
  \end{align}
\end{subequations}
The last line, Eq.~(\ref{eqn:pw_eta_s}), implicitly determines the phase
$\eta_s$ of the electromagnetic field, where the pair will be produced
on the classical trajectory.  This trajectory, as well as $\eta_s$, has
to be imaginary.  Again, this imaginary trajectory will not be the
physical trajectory of the process.  It is only the trajectory where the
exponential part of the matrix element gets stationary and hence yields
a good approximation for its value.  The part of the exponent which has
an imaginary value is given by the sum of the actions of the electron,
positron and photon:
\begin{equation}
  \Sigma W = \frac{1}{\omega_L} \left( \int^{\eta_s} p_0 \,\dd \varphi  +
  \int^{\eta_s} q_0  \, \dd \varphi -
  \int^{\eta_s} k_0 \, \dd \varphi 
  \right) \,.
  \label{eqn:remain_exp_eta}
\end{equation}
In the general case, $\eta_s$ and subsequently $\Im \Sigma W$ need to be
calculated depending on the analytic form of the given electromagnetic
field.  Usually, this has to be done numerically.  In the case of the
classical nonlinearity parameter $\xi = e E_0/m \omega_L$ being much
larger than 1, the integral can be calculated perturbatively up to order
$\mathcal{O}(1/\xi)$ and gives the same results as a constant crossed
field with $|E| = |B|$ as shown in the following.  Using the identity
\begin{equation}
  \Lambda_p = k_L p = \omega_L \left( p_0 - p_z \right) 
\end{equation}
it follows
\begin{gather}
  p_z = p_0 - \frac{\Lambda_p}{\omega_L} \,, \\
  p_z^2 = m^2 + p_x^2 + p_y^2 + p_z^2 -
  2 p_0 \frac{\Lambda_p}{\omega_L} + \frac{\Lambda_p^2}{\omega_L^2}\,,\\
  \label{eqn:p0_alt_form}
  p_0  = \frac{\omega_L}{2\Lambda_p}
  \left( m^2 + p_x^2 + p_y^2 \right) + \frac{\Lambda_p}{2 \omega_L} \,,
\end{gather}
and from Eqs.~(\ref{eqn:plane_wave_conservation}) and
(\ref{eqn:canon_kin_rel}),
\begin{equation}
  p_x(\eta) = - q_x(\eta) \quad \rightarrow \quad p_x^2(\eta) = q_x^2(\eta) \,.
  \label{eqn:px_qx_eta}
\end{equation}
Plugging Eqs.~(\ref{eqn:p0_alt_form}) and (\ref{eqn:px_qx_eta}) into
Eq.~(\ref{eqn:remain_exp_eta}) and using Eq.~(\ref{eqn:lambda_p_q_k}) gives
\begin{equation}
  \Sigma W = -\left( \frac{1}{2 \Lambda_p} + \frac{1}{2 \Lambda_q} \right) 
  \int^{\eta_s} \dd \varphi \left[ m_*^2 + p_x^2(\varphi) \right] \,.
  \label{eqn:pw_pert_exp}
\end{equation}
$\eta_s$ is implicitly fixed by Eq.~(\ref{eqn:pw_eta_s}), leading to
\begin{equation}
  p_x(\eta_s) = \pm i m_*\,,
\end{equation}
with $m_*$ being defined as in Eq.~(\ref{eqn:pxs_imag}).  Using the
four-potential Eq.~(\ref{eqn:plane_wave_potential}) and rewriting in
terms of the canonical momentum $P_x$ (constant of motion) yields
\begin{equation}
  \pm i \frac{1}{\xi} + \frac{\omega_L P_x}{e E_0} = f(\eta_s)
  \quad \text{with} \quad
  \xi = \frac{e E_0}{m_* \omega_L}\,.
  \label{eqn:f_expansion}
\end{equation}
Assuming $\xi$ being large, $\eta_s$ can be evaluated perturbatively in
powers of $1/\xi$ as
\begin{equation}
  \eta_s = 
  \eta_s^0 + \frac{1}{\xi} \eta_s^1 + \frac{1}{\xi^2} \eta_s^2 + \cdots\,.
\end{equation}
Solving Eq.~(\ref{eqn:f_expansion}) up to first order in $1/\xi$ gives
\begin{align}
  \eta_s^0 & = f^{-1}\left( \frac{\omega_L P_x}{e E_0} \right) \,, \\
  \eta_s^1 & = \pm i/f'(\eta_s^0) \,.
\end{align}
Only the part $\eta_s^0 \rightarrow \eta_s^0 + \eta_s^1/\xi$ of the
integration contour for $\Sigma W$ gives an imaginary contribution (in
$\mathcal{O}(1/\xi)$), and therefore the integral needs to be calculated just
along this path.  The integrand of $\Sigma W$ is rewritten as
\begin{equation}
  m_*^2 + p_x^2 = m_*^2 \left( 1 + \frac{p_x^2}{m_*^2} \right)
  = m_*^2 \left[ 
    1 + \left( \frac{P_x}{m_*} - \xi f(\varphi) \right)^2 
  \right] \,,
\end{equation}
and expanding $f(\varphi)$ around $\eta_s^0$,
\begin{equation}
  f(\varphi) \sim f(\eta_s^0) + f'(\eta_s^0) (\varphi - \eta_s^0) + \cdots\,,
\end{equation}
yields
\begin{equation}
  m_*^2 + p_x^2 \sim m_*^2 \left[ 1 + 
  \left[ \xi f'(\eta_s^0)(\varphi - \eta_s^2) \right]^2\right] \,.
\end{equation}
Consequently, $\Sigma W$ calculates to
\begin{multline}
   \sim  m_*^2 \int_{\eta_s^0}^{\eta_s^0 + \eta_s^1/\xi} \dd \varphi 
  \left[ 1 + \left[ \xi f'(\eta_s^0)(\varphi - \eta_s^2) \right]^2 \right] \\
  =  \pm i \frac{2}{3}\frac{m_*^2}{\xi f'(\eta_s^0)} +
  \mathcal{O}\left( \frac{1}{\xi^2} \right) \,.
\end{multline}
Taking the correct sign of $\pm i$ for exponential suppression and writing
$\xi$ explicitly gives the final result
\begin{equation}
  \label{eqn:case_pw}
  \Sigma W = W^- + W^+ = 
  \frac{i}{3} \frac{m_*^3}{\Lambda_p} \frac{\omega_L}{e E(\eta_s^0)}
  + \frac{i}{3} \frac{m_*^3}{\Lambda_q} \frac{\omega_L}{e E(\eta_s^0)} \,.
\end{equation}
Taking only the electron part $W^-$ and rewriting it as
\begin{equation}
  W^- = \frac{i}{3} \frac{m_*^3}{\omega_L (p_0 - p_z)}
  \frac{\omega_L}{e E(\eta_s^0)} =
  \frac{i}{3} \frac{m_*^3}{p_0 - p_z} \frac{1}{e E(\eta_s^0)}\,,
\end{equation}
it can be seen that this yields the same result as the constant field
approximation in Eq.~(\ref{eqn:exp_case2}) by using the identities
\begin{math}
  B = - E{k_z}/{k_0}
\end{math}
and
\begin{math}
  p^2_{0,e} - p^2_{z,e} = m^2_* 
\end{math}.  This is because the expansion only up to the order
$\mathcal{O}(1/\xi)$ corresponds to expanding the external field up to
its first derivative, giving a constant crossed field.

\bibliography{refs}

\begin{thebibliography}{55}%
\makeatletter
\providecommand \@ifxundefined [1]{%
 \@ifx{#1\undefined}
}%
\providecommand \@ifnum [1]{%
 \ifnum #1\expandafter \@firstoftwo
 \else \expandafter \@secondoftwo
 \fi
}%
\providecommand \@ifx [1]{%
 \ifx #1\expandafter \@firstoftwo
 \else \expandafter \@secondoftwo
 \fi
}%
\providecommand \natexlab [1]{#1}%
\providecommand \enquote  [1]{``#1''}%
\providecommand \bibnamefont  [1]{#1}%
\providecommand \bibfnamefont [1]{#1}%
\providecommand \citenamefont [1]{#1}%
\providecommand \href@noop [0]{\@secondoftwo}%
\providecommand \href [0]{\begingroup \@sanitize@url \@href}%
\providecommand \@href[1]{\@@startlink{#1}\@@href}%
\providecommand \@@href[1]{\endgroup#1\@@endlink}%
\providecommand \@sanitize@url [0]{\catcode `\\12\catcode `\$12\catcode
  `\&12\catcode `\#12\catcode `\^12\catcode `\_12\catcode `\%12\relax}%
\providecommand \@@startlink[1]{}%
\providecommand \@@endlink[0]{}%
\providecommand \url  [0]{\begingroup\@sanitize@url \@url }%
\providecommand \@url [1]{\endgroup\@href {#1}{\urlprefix }}%
\providecommand \urlprefix  [0]{URL }%
\providecommand \Eprint [0]{\href }%
\providecommand \doibase [0]{http://dx.doi.org/}%
\providecommand \selectlanguage [0]{\@gobble}%
\providecommand \bibinfo  [0]{\@secondoftwo}%
\providecommand \bibfield  [0]{\@secondoftwo}%
\providecommand \translation [1]{[#1]}%
\providecommand \BibitemOpen [0]{}%
\providecommand \bibitemStop [0]{}%
\providecommand \bibitemNoStop [0]{.\EOS\space}%
\providecommand \EOS [0]{\spacefactor3000\relax}%
\providecommand \BibitemShut  [1]{\csname bibitem#1\endcsname}%
\let\auto@bib@innerbib\@empty
\bibitem [{\citenamefont {Sauter}(1931)}]{sauter1931verhalten}%
  \BibitemOpen
  \bibfield  {author} {\bibinfo {author} {\bibfnamefont {F.}~\bibnamefont
  {Sauter}},\ }\href {\doibase 10.1007/BF01339461} {\bibfield  {journal}
  {\bibinfo  {journal} {Z. Phys.}\ }\textbf {\bibinfo {volume} {69}},\ \bibinfo
  {pages} {742} (\bibinfo {year} {1931})}\BibitemShut {NoStop}%
\bibitem [{\citenamefont {Heisenberg}\ and\ \citenamefont
  {Euler}(1936)}]{heisenberg1936folgerungen}%
  \BibitemOpen
  \bibfield  {author} {\bibinfo {author} {\bibfnamefont {W.}~\bibnamefont
  {Heisenberg}}\ and\ \bibinfo {author} {\bibfnamefont {H.}~\bibnamefont
  {Euler}},\ }\href {\doibase 10.1007/BF01343663} {\bibfield  {journal}
  {\bibinfo  {journal} {Z. Phys.}\ }\textbf {\bibinfo {volume} {98}},\ \bibinfo
  {pages} {714} (\bibinfo {year} {1936})}\BibitemShut {NoStop}%
\bibitem [{\citenamefont {Schwinger}(1951)}]{schwinger1951on_gauge}%
  \BibitemOpen
  \bibfield  {author} {\bibinfo {author} {\bibfnamefont {J.}~\bibnamefont
  {Schwinger}},\ }\href {\doibase 10.1103/PhysRev.82.664} {\bibfield  {journal}
  {\bibinfo  {journal} {Phys. Rev.}\ }\textbf {\bibinfo {volume} {82}},\
  \bibinfo {pages} {664} (\bibinfo {year} {1951})}\BibitemShut {NoStop}%
\bibitem [{\citenamefont {Brezin}\ and\ \citenamefont
  {Itzykson}(1970)}]{brezin1970pair}%
  \BibitemOpen
  \bibfield  {author} {\bibinfo {author} {\bibfnamefont {E.}~\bibnamefont
  {Brezin}}\ and\ \bibinfo {author} {\bibfnamefont {C.}~\bibnamefont
  {Itzykson}},\ }\href {\doibase 10.1103/PhysRevD.2.1191} {\bibfield  {journal}
  {\bibinfo  {journal} {Phys. Rev. D}\ }\textbf {\bibinfo {volume} {2}},\
  \bibinfo {pages} {1191} (\bibinfo {year} {1970})}\BibitemShut {NoStop}%
\bibitem [{\citenamefont {Ehlotzky}\ \emph {et~al.}(2009)\citenamefont
  {Ehlotzky}, \citenamefont {Krajewska},\ and\ \citenamefont
  {Kami\'nski}}]{ehlotzky2009fundamental}%
  \BibitemOpen
  \bibfield  {author} {\bibinfo {author} {\bibfnamefont {F.}~\bibnamefont
  {Ehlotzky}}, \bibinfo {author} {\bibfnamefont {K.}~\bibnamefont {Krajewska}},
  \ and\ \bibinfo {author} {\bibfnamefont {J.~Z.}\ \bibnamefont {Kami\'nski}},\
  }\href {\doibase 10.1088/0034-4885/72/4/046401} {\bibfield  {journal}
  {\bibinfo  {journal} {Rep. Prog. in Phys.}\ }\textbf {\bibinfo {volume}
  {72}},\ \bibinfo {pages} {046401} (\bibinfo {year} {2009})}\BibitemShut
  {NoStop}%
\bibitem [{\citenamefont {Ruffini}\ \emph {et~al.}(2010)\citenamefont
  {Ruffini}, \citenamefont {Vereshchagin},\ and\ \citenamefont
  {Xue}}]{ruffini2010electron}%
  \BibitemOpen
  \bibfield  {author} {\bibinfo {author} {\bibfnamefont {R.}~\bibnamefont
  {Ruffini}}, \bibinfo {author} {\bibfnamefont {G.}~\bibnamefont
  {Vereshchagin}}, \ and\ \bibinfo {author} {\bibfnamefont {S.-S.}\
  \bibnamefont {Xue}},\ }\href {\doibase 10.1016/j.physrep.2009.10.004}
  {\bibfield  {journal} {\bibinfo  {journal} {Phys. Rep.}\ }\textbf {\bibinfo
  {volume} {487}},\ \bibinfo {pages} {1} (\bibinfo {year} {2010})}\BibitemShut
  {NoStop}%
\bibitem [{\citenamefont {{Di Piazza}}\ \emph {et~al.}(2012)\citenamefont {{Di
  Piazza}}, \citenamefont {M\"uller}, \citenamefont {Hatsagortsyan},\ and\
  \citenamefont {Keitel}}]{dipiazza2012extremely}%
  \BibitemOpen
  \bibfield  {author} {\bibinfo {author} {\bibfnamefont {A.}~\bibnamefont {{Di
  Piazza}}}, \bibinfo {author} {\bibfnamefont {C.}~\bibnamefont {M\"uller}},
  \bibinfo {author} {\bibfnamefont {K.~Z.}\ \bibnamefont {Hatsagortsyan}}, \
  and\ \bibinfo {author} {\bibfnamefont {C.~H.}\ \bibnamefont {Keitel}},\
  }\href {\doibase 10.1103/RevModPhys.84.1177} {\bibfield  {journal} {\bibinfo
  {journal} {Rev. Mod. Phys.}\ }\textbf {\bibinfo {volume} {84}},\ \bibinfo
  {pages} {1177} (\bibinfo {year} {2012})}\BibitemShut {NoStop}%
\bibitem [{\citenamefont {Gies}\ and\ \citenamefont
  {Klingm\"uller}(2005)}]{gies2005pair}%
  \BibitemOpen
  \bibfield  {author} {\bibinfo {author} {\bibfnamefont {H.}~\bibnamefont
  {Gies}}\ and\ \bibinfo {author} {\bibfnamefont {K.}~\bibnamefont
  {Klingm\"uller}},\ }\href {\doibase 10.1103/PhysRevD.72.065001} {\bibfield
  {journal} {\bibinfo  {journal} {Phys. Rev. D}\ }\textbf {\bibinfo {volume}
  {72}},\ \bibinfo {pages} {065001} (\bibinfo {year} {2005})}\BibitemShut
  {NoStop}%
\bibitem [{\citenamefont {Ruf}\ \emph {et~al.}(2009)\citenamefont {Ruf},
  \citenamefont {Mocken}, \citenamefont {M\"uller}, \citenamefont
  {Hatsagortsyan},\ and\ \citenamefont {Keitel}}]{ruf2009pair}%
  \BibitemOpen
  \bibfield  {author} {\bibinfo {author} {\bibfnamefont {M.}~\bibnamefont
  {Ruf}}, \bibinfo {author} {\bibfnamefont {G.~R.}\ \bibnamefont {Mocken}},
  \bibinfo {author} {\bibfnamefont {C.}~\bibnamefont {M\"uller}}, \bibinfo
  {author} {\bibfnamefont {K.~Z.}\ \bibnamefont {Hatsagortsyan}}, \ and\
  \bibinfo {author} {\bibfnamefont {C.~H.}\ \bibnamefont {Keitel}},\ }\href
  {\doibase 10.1103/PhysRevLett.102.080402} {\bibfield  {journal} {\bibinfo
  {journal} {Phys. Rev. Lett.}\ }\textbf {\bibinfo {volume} {102}},\ \bibinfo
  {pages} {080402} (\bibinfo {year} {2009})}\BibitemShut {NoStop}%
\bibitem [{\citenamefont {Mocken}\ \emph {et~al.}(2010)\citenamefont {Mocken},
  \citenamefont {Ruf}, \citenamefont {M{\"u}ller},\ and\ \citenamefont
  {Keitel}}]{mocken2010nonperturbative}%
  \BibitemOpen
  \bibfield  {author} {\bibinfo {author} {\bibfnamefont {G.~R.}\ \bibnamefont
  {Mocken}}, \bibinfo {author} {\bibfnamefont {M.}~\bibnamefont {Ruf}},
  \bibinfo {author} {\bibfnamefont {C.}~\bibnamefont {M{\"u}ller}}, \ and\
  \bibinfo {author} {\bibfnamefont {C.~H.}\ \bibnamefont {Keitel}},\ }\href
  {\doibase 10.1103/PhysRevA.81.022122} {\bibfield  {journal} {\bibinfo
  {journal} {Phys. Rev. A}\ }\textbf {\bibinfo {volume} {81}},\ \bibinfo
  {pages} {022122} (\bibinfo {year} {2010})}\BibitemShut {NoStop}%
\bibitem [{\citenamefont {Hebenstreit}\ \emph {et~al.}(2011)\citenamefont
  {Hebenstreit}, \citenamefont {Alkofer},\ and\ \citenamefont
  {Gies}}]{hebenstreit2011particle}%
  \BibitemOpen
  \bibfield  {author} {\bibinfo {author} {\bibfnamefont {F.}~\bibnamefont
  {Hebenstreit}}, \bibinfo {author} {\bibfnamefont {R.}~\bibnamefont
  {Alkofer}}, \ and\ \bibinfo {author} {\bibfnamefont {H.}~\bibnamefont
  {Gies}},\ }\href {\doibase 10.1103/PhysRevLett.107.180403} {\bibfield
  {journal} {\bibinfo  {journal} {Phys. Rev. Lett.}\ }\textbf {\bibinfo
  {volume} {107}},\ \bibinfo {pages} {180403} (\bibinfo {year}
  {2011})}\BibitemShut {NoStop}%
\bibitem [{\citenamefont {Sokolov}\ \emph {et~al.}(2010)\citenamefont
  {Sokolov}, \citenamefont {Naumova}, \citenamefont {Nees},\ and\ \citenamefont
  {Mourou}}]{sokolov2010pair}%
  \BibitemOpen
  \bibfield  {author} {\bibinfo {author} {\bibfnamefont {I.~V.}\ \bibnamefont
  {Sokolov}}, \bibinfo {author} {\bibfnamefont {N.~M.}\ \bibnamefont
  {Naumova}}, \bibinfo {author} {\bibfnamefont {J.~A.}\ \bibnamefont {Nees}}, \
  and\ \bibinfo {author} {\bibfnamefont {G.~A.}\ \bibnamefont {Mourou}},\
  }\href {\doibase 10.1103/PhysRevLett.105.195005} {\bibfield  {journal}
  {\bibinfo  {journal} {Phys. Rev. Lett.}\ }\textbf {\bibinfo {volume} {105}},\
  \bibinfo {pages} {195005} (\bibinfo {year} {2010})}\BibitemShut {NoStop}%
\bibitem [{\citenamefont {M\"uller}\ and\ \citenamefont
  {M\"uller}(2012)}]{mueller2012polarization}%
  \BibitemOpen
  \bibfield  {author} {\bibinfo {author} {\bibfnamefont {T.-O.}\ \bibnamefont
  {M\"uller}}\ and\ \bibinfo {author} {\bibfnamefont {C.}~\bibnamefont
  {M\"uller}},\ }\href {\doibase 10.1103/PhysRevA.86.022109} {\bibfield
  {journal} {\bibinfo  {journal} {Phys. Rev. A}\ }\textbf {\bibinfo {volume}
  {86}},\ \bibinfo {pages} {022109} (\bibinfo {year} {2012})}\BibitemShut
  {NoStop}%
\bibitem [{\citenamefont {Augustin}\ and\ \citenamefont
  {M\"uller}(2013)}]{augustin2013bichromatic}%
  \BibitemOpen
  \bibfield  {author} {\bibinfo {author} {\bibfnamefont {S.}~\bibnamefont
  {Augustin}}\ and\ \bibinfo {author} {\bibfnamefont {C.}~\bibnamefont
  {M\"uller}},\ }\href {\doibase 10.1103/PhysRevA.88.022109} {\bibfield
  {journal} {\bibinfo  {journal} {Phys. Rev. A}\ }\textbf {\bibinfo {volume}
  {88}},\ \bibinfo {pages} {022109} (\bibinfo {year} {2013})}\BibitemShut
  {NoStop}%
\bibitem [{\citenamefont {Krajewska}\ \emph {et~al.}(2013)\citenamefont
  {Krajewska}, \citenamefont {M\"uller},\ and\ \citenamefont
  {Kami\ifmmode~\acute{n}\else \'{n}\fi{}ski}}]{krajewska2013betheheitler}%
  \BibitemOpen
  \bibfield  {author} {\bibinfo {author} {\bibfnamefont {K.}~\bibnamefont
  {Krajewska}}, \bibinfo {author} {\bibfnamefont {C.}~\bibnamefont {M\"uller}},
  \ and\ \bibinfo {author} {\bibfnamefont {J.~Z.}\ \bibnamefont
  {Kami\ifmmode~\acute{n}\else \'{n}\fi{}ski}},\ }\href {\doibase
  10.1103/PhysRevA.87.062107} {\bibfield  {journal} {\bibinfo  {journal} {Phys.
  Rev. A}\ }\textbf {\bibinfo {volume} {87}},\ \bibinfo {pages} {062107}
  (\bibinfo {year} {2013})}\BibitemShut {NoStop}%
\bibitem [{\citenamefont {Sch\"utzhold}\ \emph {et~al.}(2008)\citenamefont
  {Sch\"utzhold}, \citenamefont {Gies},\ and\ \citenamefont
  {Dunne}}]{schuetzhold2008dynamically}%
  \BibitemOpen
  \bibfield  {author} {\bibinfo {author} {\bibfnamefont {R.}~\bibnamefont
  {Sch\"utzhold}}, \bibinfo {author} {\bibfnamefont {H.}~\bibnamefont {Gies}},
  \ and\ \bibinfo {author} {\bibfnamefont {G.}~\bibnamefont {Dunne}},\ }\href
  {\doibase 10.1103/PhysRevLett.101.130404} {\bibfield  {journal} {\bibinfo
  {journal} {Phys. Rev. Lett.}\ }\textbf {\bibinfo {volume} {101}},\ \bibinfo
  {pages} {130404} (\bibinfo {year} {2008})}\BibitemShut {NoStop}%
\bibitem [{\citenamefont {Dunne}\ \emph {et~al.}(2009)\citenamefont {Dunne},
  \citenamefont {Gies},\ and\ \citenamefont
  {Sch\"utzhold}}]{dunne2009catalysis}%
  \BibitemOpen
  \bibfield  {author} {\bibinfo {author} {\bibfnamefont {G.~V.}\ \bibnamefont
  {Dunne}}, \bibinfo {author} {\bibfnamefont {H.}~\bibnamefont {Gies}}, \ and\
  \bibinfo {author} {\bibfnamefont {R.}~\bibnamefont {Sch\"utzhold}},\ }\href
  {\doibase 10.1103/PhysRevD.80.111301} {\bibfield  {journal} {\bibinfo
  {journal} {Phys. Rev. D}\ }\textbf {\bibinfo {volume} {80}},\ \bibinfo
  {pages} {111301} (\bibinfo {year} {2009})}\BibitemShut {NoStop}%
\bibitem [{\citenamefont {Di~Piazza}\ \emph {et~al.}(2009)\citenamefont
  {Di~Piazza}, \citenamefont {L\"otstedt}, \citenamefont {Milstein},\ and\
  \citenamefont {Keitel}}]{piazza2009barrier}%
  \BibitemOpen
  \bibfield  {author} {\bibinfo {author} {\bibfnamefont {A.}~\bibnamefont
  {Di~Piazza}}, \bibinfo {author} {\bibfnamefont {E.}~\bibnamefont
  {L\"otstedt}}, \bibinfo {author} {\bibfnamefont {A.~I.}\ \bibnamefont
  {Milstein}}, \ and\ \bibinfo {author} {\bibfnamefont {C.~H.}\ \bibnamefont
  {Keitel}},\ }\href {\doibase 10.1103/PhysRevLett.103.170403} {\bibfield
  {journal} {\bibinfo  {journal} {Phys. Rev. Lett.}\ }\textbf {\bibinfo
  {volume} {103}},\ \bibinfo {pages} {170403} (\bibinfo {year}
  {2009})}\BibitemShut {NoStop}%
\bibitem [{\citenamefont {King}\ \emph {et~al.}(2012)\citenamefont {King},
  \citenamefont {Gies},\ and\ \citenamefont {Di~Piazza}}]{king2012thermal}%
  \BibitemOpen
  \bibfield  {author} {\bibinfo {author} {\bibfnamefont {B.}~\bibnamefont
  {King}}, \bibinfo {author} {\bibfnamefont {H.}~\bibnamefont {Gies}}, \ and\
  \bibinfo {author} {\bibfnamefont {A.}~\bibnamefont {Di~Piazza}},\ }\href
  {\doibase 10.1103/PhysRevD.86.125007} {\bibfield  {journal} {\bibinfo
  {journal} {Phys. Rev. D}\ }\textbf {\bibinfo {volume} {86}},\ \bibinfo
  {pages} {125007} (\bibinfo {year} {2012})}\BibitemShut {NoStop}%
\bibitem [{\citenamefont {Krajewska}\ and\ \citenamefont
  {Kami\ifmmode~\acute{n}\else \'{n}\fi{}ski}(2012)}]{krajewska2012breit}%
  \BibitemOpen
  \bibfield  {author} {\bibinfo {author} {\bibfnamefont {K.}~\bibnamefont
  {Krajewska}}\ and\ \bibinfo {author} {\bibfnamefont {J.~Z.}\ \bibnamefont
  {Kami\ifmmode~\acute{n}\else \'{n}\fi{}ski}},\ }\href {\doibase
  10.1103/PhysRevA.86.052104} {\bibfield  {journal} {\bibinfo  {journal} {Phys.
  Rev. A}\ }\textbf {\bibinfo {volume} {86}},\ \bibinfo {pages} {052104}
  (\bibinfo {year} {2012})}\BibitemShut {NoStop}%
\bibitem [{\citenamefont {Kim}\ and\ \citenamefont
  {Page}(2006)}]{kim2006schwinger}%
  \BibitemOpen
  \bibfield  {author} {\bibinfo {author} {\bibfnamefont {S.~P.}\ \bibnamefont
  {Kim}}\ and\ \bibinfo {author} {\bibfnamefont {D.~N.}\ \bibnamefont {Page}},\
  }\href {\doibase 10.1103/PhysRevD.73.065020} {\bibfield  {journal} {\bibinfo
  {journal} {Phys. Rev. D}\ }\textbf {\bibinfo {volume} {73}},\ \bibinfo
  {pages} {065020} (\bibinfo {year} {2006})}\BibitemShut {NoStop}%
\bibitem [{\citenamefont {Su}\ \emph {et~al.}(2012{\natexlab{a}})\citenamefont
  {Su}, \citenamefont {Su}, \citenamefont {Lv}, \citenamefont {Jiang},
  \citenamefont {Lu}, \citenamefont {Sheng},\ and\ \citenamefont
  {Grobe}}]{su2012magnetic}%
  \BibitemOpen
  \bibfield  {author} {\bibinfo {author} {\bibfnamefont {Q.}~\bibnamefont
  {Su}}, \bibinfo {author} {\bibfnamefont {W.}~\bibnamefont {Su}}, \bibinfo
  {author} {\bibfnamefont {Q.~Z.}\ \bibnamefont {Lv}}, \bibinfo {author}
  {\bibfnamefont {M.}~\bibnamefont {Jiang}}, \bibinfo {author} {\bibfnamefont
  {X.}~\bibnamefont {Lu}}, \bibinfo {author} {\bibfnamefont {Z.~M.}\
  \bibnamefont {Sheng}}, \ and\ \bibinfo {author} {\bibfnamefont
  {R.}~\bibnamefont {Grobe}},\ }\href {\doibase 10.1103/PhysRevLett.109.253202}
  {\bibfield  {journal} {\bibinfo  {journal} {Phys. Rev. Lett.}\ }\textbf
  {\bibinfo {volume} {109}},\ \bibinfo {pages} {253202} (\bibinfo {year}
  {2012}{\natexlab{a}})}\BibitemShut {NoStop}%
\bibitem [{\citenamefont {Su}\ \emph {et~al.}(2012{\natexlab{b}})\citenamefont
  {Su}, \citenamefont {Jiang}, \citenamefont {Lv}, \citenamefont {Li},
  \citenamefont {Sheng}, \citenamefont {Grobe},\ and\ \citenamefont
  {Su}}]{su2012suppression}%
  \BibitemOpen
  \bibfield  {author} {\bibinfo {author} {\bibfnamefont {W.}~\bibnamefont
  {Su}}, \bibinfo {author} {\bibfnamefont {M.}~\bibnamefont {Jiang}}, \bibinfo
  {author} {\bibfnamefont {Z.~Q.}\ \bibnamefont {Lv}}, \bibinfo {author}
  {\bibfnamefont {Y.~J.}\ \bibnamefont {Li}}, \bibinfo {author} {\bibfnamefont
  {Z.~M.}\ \bibnamefont {Sheng}}, \bibinfo {author} {\bibfnamefont
  {R.}~\bibnamefont {Grobe}}, \ and\ \bibinfo {author} {\bibfnamefont
  {Q.}~\bibnamefont {Su}},\ }\href {\doibase 10.1103/physreva.86.013422}
  {\bibfield  {journal} {\bibinfo  {journal} {Phys. Rev. A}\ }\textbf {\bibinfo
  {volume} {86}},\ \bibinfo {pages} {013422} (\bibinfo {year}
  {2012}{\natexlab{b}})}\BibitemShut {NoStop}%
\bibitem [{\citenamefont {Lv}\ \emph {et~al.}(2013)\citenamefont {Lv},
  \citenamefont {Li}, \citenamefont {Grobe},\ and\ \citenamefont
  {Su}}]{lv2013pairtunnel}%
  \BibitemOpen
  \bibfield  {author} {\bibinfo {author} {\bibfnamefont {Q.~Z.}\ \bibnamefont
  {Lv}}, \bibinfo {author} {\bibfnamefont {Y.~J.}\ \bibnamefont {Li}}, \bibinfo
  {author} {\bibfnamefont {R.}~\bibnamefont {Grobe}}, \ and\ \bibinfo {author}
  {\bibfnamefont {Q.}~\bibnamefont {Su}},\ }\href {\doibase
  10.1103/PhysRevA.88.033403} {\bibfield  {journal} {\bibinfo  {journal} {Phys.
  Rev. A}\ }\textbf {\bibinfo {volume} {88}},\ \bibinfo {pages} {033403}
  (\bibinfo {year} {2013})}\BibitemShut {NoStop}%
\bibitem [{\citenamefont {Hebenstreit}\ \emph {et~al.}(2013)\citenamefont
  {Hebenstreit}, \citenamefont {Berges},\ and\ \citenamefont
  {Gelfand}}]{hebenstreit2013one}%
  \BibitemOpen
  \bibfield  {author} {\bibinfo {author} {\bibfnamefont {F.}~\bibnamefont
  {Hebenstreit}}, \bibinfo {author} {\bibfnamefont {J.}~\bibnamefont {Berges}},
  \ and\ \bibinfo {author} {\bibfnamefont {D.}~\bibnamefont {Gelfand}},\ }\href
  {\doibase 10.1103/PhysRevD.87.105006} {\bibfield  {journal} {\bibinfo
  {journal} {Phys. Rev. D}\ }\textbf {\bibinfo {volume} {87}},\ \bibinfo
  {pages} {105006} (\bibinfo {year} {2013})}\BibitemShut {NoStop}%
\bibitem [{\citenamefont {Liu}\ \emph {et~al.}(2014)\citenamefont {Liu},
  \citenamefont {Jiang}, \citenamefont {Lv}, \citenamefont {Li}, \citenamefont
  {Grobe},\ and\ \citenamefont {Su}}]{liu2014population}%
  \BibitemOpen
  \bibfield  {author} {\bibinfo {author} {\bibfnamefont {Y.}~\bibnamefont
  {Liu}}, \bibinfo {author} {\bibfnamefont {M.}~\bibnamefont {Jiang}}, \bibinfo
  {author} {\bibfnamefont {Q.~Z.}\ \bibnamefont {Lv}}, \bibinfo {author}
  {\bibfnamefont {Y.~T.}\ \bibnamefont {Li}}, \bibinfo {author} {\bibfnamefont
  {R.}~\bibnamefont {Grobe}}, \ and\ \bibinfo {author} {\bibfnamefont
  {Q.}~\bibnamefont {Su}},\ }\href {\doibase 10.1103/PhysRevA.89.012127}
  {\bibfield  {journal} {\bibinfo  {journal} {Phys. Rev. A}\ }\textbf {\bibinfo
  {volume} {89}},\ \bibinfo {pages} {012127} (\bibinfo {year}
  {2014})}\BibitemShut {NoStop}%
\bibitem [{\citenamefont {Hebenstreit}\ and\ \citenamefont
  {Berges}(2014)}]{hebenstreit2014massless}%
  \BibitemOpen
  \bibfield  {author} {\bibinfo {author} {\bibfnamefont {F.}~\bibnamefont
  {Hebenstreit}}\ and\ \bibinfo {author} {\bibfnamefont {J.}~\bibnamefont
  {Berges}},\ }\href {\doibase 10.1103/PhysRevD.90.045034} {\bibfield
  {journal} {\bibinfo  {journal} {Phys. Rev. D}\ }\textbf {\bibinfo {volume}
  {90}},\ \bibinfo {pages} {045034} (\bibinfo {year} {2014})}\BibitemShut
  {NoStop}%
\bibitem [{\citenamefont {Jiang}\ \emph {et~al.}(2014)\citenamefont {Jiang},
  \citenamefont {Lv}, \citenamefont {Liu}, \citenamefont {Grobe},\ and\
  \citenamefont {Su}}]{jiang2014pair}%
  \BibitemOpen
  \bibfield  {author} {\bibinfo {author} {\bibfnamefont {M.}~\bibnamefont
  {Jiang}}, \bibinfo {author} {\bibfnamefont {Q.~Z.}\ \bibnamefont {Lv}},
  \bibinfo {author} {\bibfnamefont {Y.}~\bibnamefont {Liu}}, \bibinfo {author}
  {\bibfnamefont {R.}~\bibnamefont {Grobe}}, \ and\ \bibinfo {author}
  {\bibfnamefont {Q.}~\bibnamefont {Su}},\ }\href {\doibase
  10.1103/PhysRevA.90.032101} {\bibfield  {journal} {\bibinfo  {journal} {Phys.
  Rev. A}\ }\textbf {\bibinfo {volume} {90}},\ \bibinfo {pages} {032101}
  (\bibinfo {year} {2014})}\BibitemShut {NoStop}%
\bibitem [{\citenamefont {Kohlf\"urst}\ \emph {et~al.}(2014)\citenamefont
  {Kohlf\"urst}, \citenamefont {Gies},\ and\ \citenamefont
  {Alkofer}}]{kohlfurst2014effective}%
  \BibitemOpen
  \bibfield  {author} {\bibinfo {author} {\bibfnamefont {C.}~\bibnamefont
  {Kohlf\"urst}}, \bibinfo {author} {\bibfnamefont {H.}~\bibnamefont {Gies}}, \
  and\ \bibinfo {author} {\bibfnamefont {R.}~\bibnamefont {Alkofer}},\ }\href
  {\doibase 10.1103/PhysRevLett.112.050402} {\bibfield  {journal} {\bibinfo
  {journal} {Phys. Rev. Lett.}\ }\textbf {\bibinfo {volume} {112}},\ \bibinfo
  {pages} {050402} (\bibinfo {year} {2014})}\BibitemShut {NoStop}%
\bibitem [{\citenamefont {Mourou}\ \emph {et~al.}(2006)\citenamefont {Mourou},
  \citenamefont {Tajima},\ and\ \citenamefont {Bulanov}}]{mourou2006optics}%
  \BibitemOpen
  \bibfield  {author} {\bibinfo {author} {\bibfnamefont {G.~A.}\ \bibnamefont
  {Mourou}}, \bibinfo {author} {\bibfnamefont {T.}~\bibnamefont {Tajima}}, \
  and\ \bibinfo {author} {\bibfnamefont {S.~V.}\ \bibnamefont {Bulanov}},\
  }\href {\doibase 10.1103/RevModPhys.78.309} {\bibfield  {journal} {\bibinfo
  {journal} {Rev. Mod. Phys.}\ }\textbf {\bibinfo {volume} {78}},\ \bibinfo
  {pages} {309} (\bibinfo {year} {2006})}\BibitemShut {NoStop}%
\bibitem [{\citenamefont {Mourou}\ \emph {et~al.}(2012)\citenamefont {Mourou},
  \citenamefont {Fisch}, \citenamefont {Malkin}, \citenamefont {Toroker},
  \citenamefont {Khazanov}, \citenamefont {Sergeev}, \citenamefont {Tajima},\
  and\ \citenamefont {Le~Garrec}}]{mourou2012exawatt}%
  \BibitemOpen
  \bibfield  {author} {\bibinfo {author} {\bibfnamefont {G.~A.}\ \bibnamefont
  {Mourou}}, \bibinfo {author} {\bibfnamefont {N.~J.}\ \bibnamefont {Fisch}},
  \bibinfo {author} {\bibfnamefont {V.~M.}\ \bibnamefont {Malkin}}, \bibinfo
  {author} {\bibfnamefont {Z.}~\bibnamefont {Toroker}}, \bibinfo {author}
  {\bibfnamefont {E.~A.}\ \bibnamefont {Khazanov}}, \bibinfo {author}
  {\bibfnamefont {A.~M.}\ \bibnamefont {Sergeev}}, \bibinfo {author}
  {\bibfnamefont {T.}~\bibnamefont {Tajima}}, \ and\ \bibinfo {author}
  {\bibfnamefont {B.}~\bibnamefont {Le~Garrec}},\ }\href {\doibase
  10.1016/j.optcom.2011.10.089} {\bibfield  {journal} {\bibinfo  {journal}
  {Opt. Commun.}\ }\textbf {\bibinfo {volume} {285}},\ \bibinfo {pages} {720}
  (\bibinfo {year} {2012})}\BibitemShut {NoStop}%
\bibitem [{\citenamefont {Szpak}\ and\ \citenamefont
  {Sch{\"u}tzhold}(2012)}]{szpak2012optical_lattice}%
  \BibitemOpen
  \bibfield  {author} {\bibinfo {author} {\bibfnamefont {N.}~\bibnamefont
  {Szpak}}\ and\ \bibinfo {author} {\bibfnamefont {R.}~\bibnamefont
  {Sch{\"u}tzhold}},\ }\href {\doibase 10.1088/1367-2630/14/3/035001}
  {\bibfield  {journal} {\bibinfo  {journal} {New J. Phys.}\ }\textbf {\bibinfo
  {volume} {14}},\ \bibinfo {pages} {035001} (\bibinfo {year}
  {2012})}\BibitemShut {NoStop}%
\bibitem [{\citenamefont {Fedotov}\ \emph {et~al.}(2010)\citenamefont
  {Fedotov}, \citenamefont {Narozhny}, \citenamefont {Mourou},\ and\
  \citenamefont {Korn}}]{fedotov2010limitations}%
  \BibitemOpen
  \bibfield  {author} {\bibinfo {author} {\bibfnamefont {A.~M.}\ \bibnamefont
  {Fedotov}}, \bibinfo {author} {\bibfnamefont {N.~B.}\ \bibnamefont
  {Narozhny}}, \bibinfo {author} {\bibfnamefont {G.}~\bibnamefont {Mourou}}, \
  and\ \bibinfo {author} {\bibfnamefont {G.}~\bibnamefont {Korn}},\ }\href
  {\doibase 10.1103/PhysRevLett.105.080402} {\bibfield  {journal} {\bibinfo
  {journal} {Phys. Rev. Lett.}\ }\textbf {\bibinfo {volume} {105}},\ \bibinfo
  {pages} {080402} (\bibinfo {year} {2010})}\BibitemShut {NoStop}%
\bibitem [{\citenamefont {Popov}(1972)}]{popov1972pair}%
  \BibitemOpen
  \bibfield  {author} {\bibinfo {author} {\bibfnamefont {V.~S.}\ \bibnamefont
  {Popov}},\ }\href@noop {} {\bibfield  {journal} {\bibinfo  {journal} {Sov.
  Phys. JETP}\ }\textbf {\bibinfo {volume} {34}},\ \bibinfo {pages} {709}
  (\bibinfo {year} {1972})}\BibitemShut {NoStop}%
\bibitem [{\citenamefont {Keldysh}(1965)}]{keldysh1965ionization}%
  \BibitemOpen
  \bibfield  {author} {\bibinfo {author} {\bibfnamefont {L.~V.}\ \bibnamefont
  {Keldysh}},\ }\href {http://jetp.ac.ru/cgi-bin/e/index/e/20/5/p1307?a=list}
  {\bibfield  {journal} {\bibinfo  {journal} {Sov. Phys. JETP}\ }\textbf
  {\bibinfo {volume} {20}},\ \bibinfo {pages} {1307} (\bibinfo {year}
  {1965})}\BibitemShut {NoStop}%
\bibitem [{\citenamefont {Popov}(1973)}]{popov1973imaginary_time}%
  \BibitemOpen
  \bibfield  {author} {\bibinfo {author} {\bibfnamefont {V.~S.}\ \bibnamefont
  {Popov}},\ }\href {http://jetp.ac.ru/cgi-bin/e/index/e/36/5/p840?a=list}
  {\bibfield  {journal} {\bibinfo  {journal} {Sov. Phys. JETP}\ }\textbf
  {\bibinfo {volume} {36}},\ \bibinfo {pages} {840} (\bibinfo {year}
  {1973})}\BibitemShut {NoStop}%
\bibitem [{\citenamefont {Popov}(2005)}]{popov2005imaginary}%
  \BibitemOpen
  \bibfield  {author} {\bibinfo {author} {\bibfnamefont {V.~S.}\ \bibnamefont
  {Popov}},\ }\href {\doibase 10.1134/1.1903097} {\bibfield  {journal}
  {\bibinfo  {journal} {Phys. Atom. Nucl.}\ }\textbf {\bibinfo {volume} {68}},\
  \bibinfo {pages} {686} (\bibinfo {year} {2005})}\BibitemShut {NoStop}%
\bibitem [{\citenamefont {Popruzhenko}\ \emph {et~al.}(2008)\citenamefont
  {Popruzhenko}, \citenamefont {Mur}, \citenamefont {Popov},\ and\
  \citenamefont {Bauer}}]{popru2008strong}%
  \BibitemOpen
  \bibfield  {author} {\bibinfo {author} {\bibfnamefont {S.~V.}\ \bibnamefont
  {Popruzhenko}}, \bibinfo {author} {\bibfnamefont {V.~D.}\ \bibnamefont
  {Mur}}, \bibinfo {author} {\bibfnamefont {V.~S.}\ \bibnamefont {Popov}}, \
  and\ \bibinfo {author} {\bibfnamefont {D.}~\bibnamefont {Bauer}},\ }\href
  {\doibase 10.1103/PhysRevLett.101.193003} {\bibfield  {journal} {\bibinfo
  {journal} {Phys. Rev. Lett.}\ }\textbf {\bibinfo {volume} {101}},\ \bibinfo
  {pages} {193003} (\bibinfo {year} {2008})}\BibitemShut {NoStop}%
\bibitem [{\citenamefont {Popruzhenko}\ \emph {et~al.}(2009)\citenamefont
  {Popruzhenko}, \citenamefont {Mur}, \citenamefont {Popov},\ and\
  \citenamefont {Bauer}}]{popru2009multi}%
  \BibitemOpen
  \bibfield  {author} {\bibinfo {author} {\bibfnamefont {S.}~\bibnamefont
  {Popruzhenko}}, \bibinfo {author} {\bibfnamefont {V.}~\bibnamefont {Mur}},
  \bibinfo {author} {\bibfnamefont {V.}~\bibnamefont {Popov}}, \ and\ \bibinfo
  {author} {\bibfnamefont {D.}~\bibnamefont {Bauer}},\ }\href {\doibase
  10.1134/S1063776109060053} {\bibfield  {journal} {\bibinfo  {journal} {Sov.
  Phys. JETP}\ }\textbf {\bibinfo {volume} {108}},\ \bibinfo {pages} {947}
  (\bibinfo {year} {2009})}\BibitemShut {NoStop}%
\bibitem [{\citenamefont {{Casta{\~n}eda Cort{\'e}s}}\ \emph
  {et~al.}(2011)\citenamefont {{Casta{\~n}eda Cort{\'e}s}}, \citenamefont
  {Popruzhenko}, \citenamefont {Bauer},\ and\ \citenamefont
  {P{\'a}lffy}}]{cortes2011laser}%
  \BibitemOpen
  \bibfield  {author} {\bibinfo {author} {\bibfnamefont {H.~M.}\ \bibnamefont
  {{Casta{\~n}eda Cort{\'e}s}}}, \bibinfo {author} {\bibfnamefont {S.~V.}\
  \bibnamefont {Popruzhenko}}, \bibinfo {author} {\bibfnamefont
  {D.}~\bibnamefont {Bauer}}, \ and\ \bibinfo {author} {\bibfnamefont
  {A.}~\bibnamefont {P{\'a}lffy}},\ }\href {\doibase
  10.1088/1367-2630/13/6/063007} {\bibfield  {journal} {\bibinfo  {journal}
  {New J. Phys.}\ }\textbf {\bibinfo {volume} {13}},\ \bibinfo {pages} {063007}
  (\bibinfo {year} {2011})}\BibitemShut {NoStop}%
\bibitem [{\citenamefont {Reiss}(2008)}]{reiss2008limits}%
  \BibitemOpen
  \bibfield  {author} {\bibinfo {author} {\bibfnamefont {H.~R.}\ \bibnamefont
  {Reiss}},\ }\href {\doibase 10.1103/PhysRevLett.101.043002} {\bibfield
  {journal} {\bibinfo  {journal} {Phys. Rev. Lett.}\ }\textbf {\bibinfo
  {volume} {101}},\ \bibinfo {pages} {043002} (\bibinfo {year}
  {2008})}\BibitemShut {NoStop}%
\bibitem [{\citenamefont {Reiss}(2014)}]{reiss2014tunneling}%
  \BibitemOpen
  \bibfield  {author} {\bibinfo {author} {\bibfnamefont {H.~R.}\ \bibnamefont
  {Reiss}},\ }\href {\doibase 10.1088/0953-4075/47/20/204006} {\bibfield
  {journal} {\bibinfo  {journal} {J. Phys. B: At., Mol. Opt. Phys.}\ }\textbf
  {\bibinfo {volume} {47}},\ \bibinfo {pages} {204006} (\bibinfo {year}
  {2014})}\BibitemShut {NoStop}%
\bibitem [{\citenamefont {Klaiber}\ \emph {et~al.}(2013)\citenamefont
  {Klaiber}, \citenamefont {Yakaboylu}, \citenamefont {Bauke}, \citenamefont
  {Hatsagortsyan},\ and\ \citenamefont {Keitel}}]{klaiber2013under}%
  \BibitemOpen
  \bibfield  {author} {\bibinfo {author} {\bibfnamefont {M.}~\bibnamefont
  {Klaiber}}, \bibinfo {author} {\bibfnamefont {E.}~\bibnamefont {Yakaboylu}},
  \bibinfo {author} {\bibfnamefont {H.}~\bibnamefont {Bauke}}, \bibinfo
  {author} {\bibfnamefont {K.~Z.}\ \bibnamefont {Hatsagortsyan}}, \ and\
  \bibinfo {author} {\bibfnamefont {C.~H.}\ \bibnamefont {Keitel}},\ }\href
  {\doibase 10.1103/PhysRevLett.110.153004} {\bibfield  {journal} {\bibinfo
  {journal} {Phys. Rev. Lett.}\ }\textbf {\bibinfo {volume} {110}},\ \bibinfo
  {pages} {153004} (\bibinfo {year} {2013})}\BibitemShut {NoStop}%
\bibitem [{\citenamefont {Yakaboylu}\ \emph {et~al.}(2013)\citenamefont
  {Yakaboylu}, \citenamefont {Klaiber}, \citenamefont {Bauke}, \citenamefont
  {Hatsagortsyan},\ and\ \citenamefont {Keitel}}]{yakaboylu2013relativistic}%
  \BibitemOpen
  \bibfield  {author} {\bibinfo {author} {\bibfnamefont {E.}~\bibnamefont
  {Yakaboylu}}, \bibinfo {author} {\bibfnamefont {M.}~\bibnamefont {Klaiber}},
  \bibinfo {author} {\bibfnamefont {H.}~\bibnamefont {Bauke}}, \bibinfo
  {author} {\bibfnamefont {K.~Z.}\ \bibnamefont {Hatsagortsyan}}, \ and\
  \bibinfo {author} {\bibfnamefont {C.~H.}\ \bibnamefont {Keitel}},\ }\href
  {\doibase 10.1103/PhysRevA.88.063421} {\bibfield  {journal} {\bibinfo
  {journal} {Phys. Rev. A}\ }\textbf {\bibinfo {volume} {88}},\ \bibinfo
  {pages} {063421} (\bibinfo {year} {2013})}\BibitemShut {NoStop}%
\bibitem [{\citenamefont {Gitman}\ \emph {et~al.}(1991)\citenamefont {Gitman},
  \citenamefont {Fradkin},\ and\ \citenamefont
  {Shvartsman}}]{gitman1991quantum}%
  \BibitemOpen
  \bibfield  {author} {\bibinfo {author} {\bibfnamefont {D.~M.}\ \bibnamefont
  {Gitman}}, \bibinfo {author} {\bibfnamefont {E.~S.}\ \bibnamefont {Fradkin}},
  \ and\ \bibinfo {author} {\bibfnamefont {S.~M.}\ \bibnamefont {Shvartsman}},\
  }\href@noop {} {\emph {\bibinfo {title} {Quantum Electrodynamics with
  Unstable Vacuum}}},\ Springer Series in Nuclear and Particle Physics\
  (\bibinfo  {publisher} {Springer},\ \bibinfo {address} {Berlin},\ \bibinfo
  {year} {1991})\BibitemShut {NoStop}%
\bibitem [{\citenamefont {Furry}(1951)}]{furry1951onbound}%
  \BibitemOpen
  \bibfield  {author} {\bibinfo {author} {\bibfnamefont {W.~H.}\ \bibnamefont
  {Furry}},\ }\href {\doibase 10.1103/PhysRev.81.115} {\bibfield  {journal}
  {\bibinfo  {journal} {Phys. Rev.}\ }\textbf {\bibinfo {volume} {81}},\
  \bibinfo {pages} {115} (\bibinfo {year} {1951})}\BibitemShut {NoStop}%
\bibitem [{Note3()}]{Note3}%
  \BibitemOpen
  \bibinfo {note} {This can be shown best in the gauge $A^\mu = (-E x, 0, 0, -B
  x)$ for the general constant crossed field case.}\BibitemShut {Stop}%
\bibitem [{\citenamefont {Nikishov}(1970)}]{nikishov1970pair}%
  \BibitemOpen
  \bibfield  {author} {\bibinfo {author} {\bibfnamefont {A.~I.}\ \bibnamefont
  {Nikishov}},\ }\href
  {http://www.jetp.ac.ru/cgi-bin/e/index/e/30/4/p660?a=list} {\bibfield
  {journal} {\bibinfo  {journal} {Sov. Phys. JETP}\ }\textbf {\bibinfo {volume}
  {30}},\ \bibinfo {pages} {660} (\bibinfo {year} {1970})}\BibitemShut
  {NoStop}%
\bibitem [{\citenamefont {Reiss}(1962)}]{reiss1962absorption}%
  \BibitemOpen
  \bibfield  {author} {\bibinfo {author} {\bibfnamefont {H.~R.}\ \bibnamefont
  {Reiss}},\ }\href {\doibase 10.1063/1.1703787} {\bibfield  {journal}
  {\bibinfo  {journal} {J. Math. Phys.}\ }\textbf {\bibinfo {volume} {3}},\
  \bibinfo {pages} {59} (\bibinfo {year} {1962})}\BibitemShut {NoStop}%
\bibitem [{\citenamefont {Nikishov}(1985)}]{nikishov1985problems}%
  \BibitemOpen
  \bibfield  {author} {\bibinfo {author} {\bibfnamefont {A.~I.}\ \bibnamefont
  {Nikishov}},\ }\href {\doibase 10.1007/BF01120143} {\bibfield  {journal}
  {\bibinfo  {journal} {J. Russ. Laser Res.}\ }\textbf {\bibinfo {volume}
  {6}},\ \bibinfo {pages} {619} (\bibinfo {year} {1985})}\BibitemShut {NoStop}%
\bibitem [{\citenamefont {Popov}\ \emph {et~al.}(1968)\citenamefont {Popov},
  \citenamefont {Kuznetsov},\ and\ \citenamefont
  {Perelomov}}]{popov1968quasiclassical}%
  \BibitemOpen
  \bibfield  {author} {\bibinfo {author} {\bibfnamefont {V.~S.}\ \bibnamefont
  {Popov}}, \bibinfo {author} {\bibfnamefont {V.~P.}\ \bibnamefont
  {Kuznetsov}}, \ and\ \bibinfo {author} {\bibfnamefont {A.~M.}\ \bibnamefont
  {Perelomov}},\ }\href {http://jetp.ac.ru/cgi-bin/e/index/e/26/1/p222?a=list}
  {\bibfield  {journal} {\bibinfo  {journal} {Sov. Phys. JETP}\ }\textbf
  {\bibinfo {volume} {26}},\ \bibinfo {pages} {222} (\bibinfo {year}
  {1968})}\BibitemShut {NoStop}%
\bibitem [{\citenamefont {Marinov}\ and\ \citenamefont
  {Popov}(1972)}]{marinov1972pair}%
  \BibitemOpen
  \bibfield  {author} {\bibinfo {author} {\bibfnamefont {M.~S.}\ \bibnamefont
  {Marinov}}\ and\ \bibinfo {author} {\bibfnamefont {V.~S.}\ \bibnamefont
  {Popov}},\ }\href@noop {} {\bibfield  {journal} {\bibinfo  {journal} {Sov. J.
  Nucl. Phys.}\ }\textbf {\bibinfo {volume} {15}},\ \bibinfo {pages} {702}
  (\bibinfo {year} {1972})}\BibitemShut {NoStop}%
\bibitem [{\citenamefont {Pauli}(1932)}]{pauli1932diracs}%
  \BibitemOpen
  \bibfield  {author} {\bibinfo {author} {\bibfnamefont {W.}~\bibnamefont
  {Pauli}},\ }\href {\doibase 10.5169/seals-110167} {\bibfield  {journal}
  {\bibinfo  {journal} {Helv. Phys. Acta}\ }\textbf {\bibinfo {volume} {5}},\
  \bibinfo {pages} {179} (\bibinfo {year} {1932})}\BibitemShut {NoStop}%
\bibitem [{Note4()}]{Note4}%
  \BibitemOpen
  \bibinfo {note} {As we do not consider spin effects in this paper, the given
  actions do not include any spin or polarization terms.}\BibitemShut {Stop}%
\bibitem [{Note5()}]{Note5}%
  \BibitemOpen
  \bibinfo {note} {Using usual quantization rules, the photon will have no
  gauge, and therefore $K = k$. Furthermore, $P = p - e A$ and $Q = q + e A$,
  which gives $ P + Q - K = p + q -k$.}\BibitemShut {Stop}%
\end{thebibliography}%

\end{document}